\documentclass[a4paper,fleqn,usenatbib]{mnras}
\usepackage{amsmath}
\usepackage{epsfig} 
\usepackage{graphicx}	
\usepackage{amssymb}	
\usepackage{xcolor}

\newcommand{\be}{\begin{equation}}
\newcommand{\e}{\end{equation}}
\newcommand{\bear}{\begin{eqnarray}}
\newcommand{\ear}{\end{eqnarray}}

\def\aj{AJ}
\def\apj{ApJ}
\def\apjs{ApJS}
\def\jcap{JCAP}
\def\mnras{MNRAS}
\def\mnrasl{MNRAS Letters}
\def\aap{A\&A}
\def\prd{Physical Review D}
      
\def\apjs{ApJS}
\def\apjl{ApJ Letters}

\title{} \author{} \title[Equation of state parameters from entropy] {Can we
  constrain the dark energy equation of state parameters using
  configuration entropy?}
    
\author[Das, B. and Pandey, B.] {Biswajit
  Das\thanks{E-mail:bishoophy@gmail.com} and {Biswajit
    Pandey\thanks{E-mail: biswap@visva-bharati.ac.in}} \\ Department of
    Physics, Visva-Bharati University, Santiniketan, Birbhum, 731235,
    India\\ }
   
 \date{\today}

 \pubyear{2019}
  \begin{document}
\label{firstpage}
\pagerange{\pageref{firstpage}--\pageref{lastpage}} 

\maketitle
\begin{abstract}
We propose a new scheme for constraining the dark energy equation of
state parameter/parameters based on the study of the evolution of the
configuration entropy. We analyze a set of one parameter and two
parameter dynamical dark energy models and find that the derivative of
the configuration entropy in all the dynamical dark energy models
exhibit a minimum. The magnitude of the minimum of the entropy rate is
decided by both the parametrization of the equation of state as well
as the associated parameters. The location of the minimum of the
entropy rate is less sensitive to the form of the parametrization but
depends on the associated parameters. We determine the best fit
equations for the location and magnitude of the minimum of the entropy
rate in terms of the parameter/parameters of the dark energy equation
of state. These relations would allow us to constrain the dark energy
equation of state parameter/parameters for any given parametrization
provided the evolution of the configuration entropy in the Universe is
known from observations.
\end{abstract}
\begin{keywords}
         methods: analytical - cosmology: theory - large scale
         structure of the Universe.
       \end{keywords}

\section {Introduction}
 The observations \citep{riess, perlmutter} tell us that the Universe
 is currently undergoing an accelerated expansion which remains one of
 the unsolved mysteries in modern cosmology. The accelerated expansion
 is very often explained by invoking a hypothetical component called
 dark energy. The dark energy is believed to have a negative pressure
 which drives the cosmic acceleration despite the presence of matter
 in the Universe and the attractive nature of gravity.

 The simplest candidate for dark energy is the cosmological constant
 which was originally introduced by Einstein in his General Theory of
 Relativity to achieve a stationary Universe. This hypothetical
 component has a constant energy density throughout the entire
 history of the Universe and has become the most dominant component
 only in the recent past. The origin of the cosmological constant is
 often linked to the vacuum energy. But unfortunately the theoretical
 value of the vacuum energy predicted by quantum field theory is
 $10^{120}$ times larger than the tiny observed value of the
 cosmological constant. This huge discrepancy points out that we still
 lack a complete theoretical understanding of the nature and origin of
 the cosmological constant.

There are other alternative models of dark energy like quintessence
\citep{ratra, caldwell} and k-essence \citep{armendariz} which are
based on the modifications of the matter side of the Einstein's field
equations. A number of alternatives such as $f(R)$ gravity
\citep{buchdahl} and scalar tensor theories \citep{bransdicke} have
been introduced by modifying the geometric side of the Einstein's
field equations. A detailed discussion on these dark energy models can
be found in \citet{copeland} and \citet{de2010}. Besides these, a
number of other interesting proposals originating from different
physically motivated ideas include the backreaction mechanism
\citep{buchert2k}, effect of a large local void \citep{tomita01,
  hunt}, entropic force \citep{easson}, extra-dimension
\citep{milton}, entropy maximization \citep{radicella, pavon1},
information storage in the spacetime \citep{paddy, paddyhamsa} and
configuration entropy of the Universe \citep{pandey1, pandey3}.

The possibility of a dynamical dark energy \citep{ratra, caldwell,
  armendariz} is a logically consistent alternative to the
cosmological constant which can be constrained by observations.  The
phenomenological approach toward this is to introduce an equation of
state (EoS) which is not constant in time. This is a generic approach
and any assumption of the underlying scalar field and its dynamics is
reflected in the equation of state. Many such parametrizations have
been proposed in the literature. The value of the parameters in these
parametrizations are constrained from different observational datasets
such as SNIa, CMB, BAO.

  \citet{pandey1} propose that the transition of the Universe from a
  highly uniform and smooth state to a highly irregular and clumpy
  state would lead to a gradual dissipation of the configuration
  entropy of the mass distribution in the Universe. The evolution of
  the configuration entropy depends on the growth rate of structure
  formation in the Universe and hence can be used to distinguish
  different models from each other.  \citet{das} consider a set of two
  parameter models of dynamical dark energy and show that the
  evolution of the configuration entropy may help us to distinguish
  the different dark energy parametrizations. In a recent work,
  \citet{pandey2} show that the second derivative of the configuration
  entropy exhibits a prominent peak at the $\Lambda$-matter equality
  which can be used to constrain the values of the matter density and
  the cosmological constant.
  
  In the present work, we consider a number of one parameter and two
  parameter models of dynamical dark energy along with the
  $\Lambda$CDM model and study the entropy rate in these models. We
  analyze the dependence of the entropy rate on the
  parameter/parameters associated with the dark energy equation of
  state and propose a new scheme to constrain them from future
  observations.

\begin{figure*}
  \resizebox{7 cm}{!}{\rotatebox{0}{\includegraphics{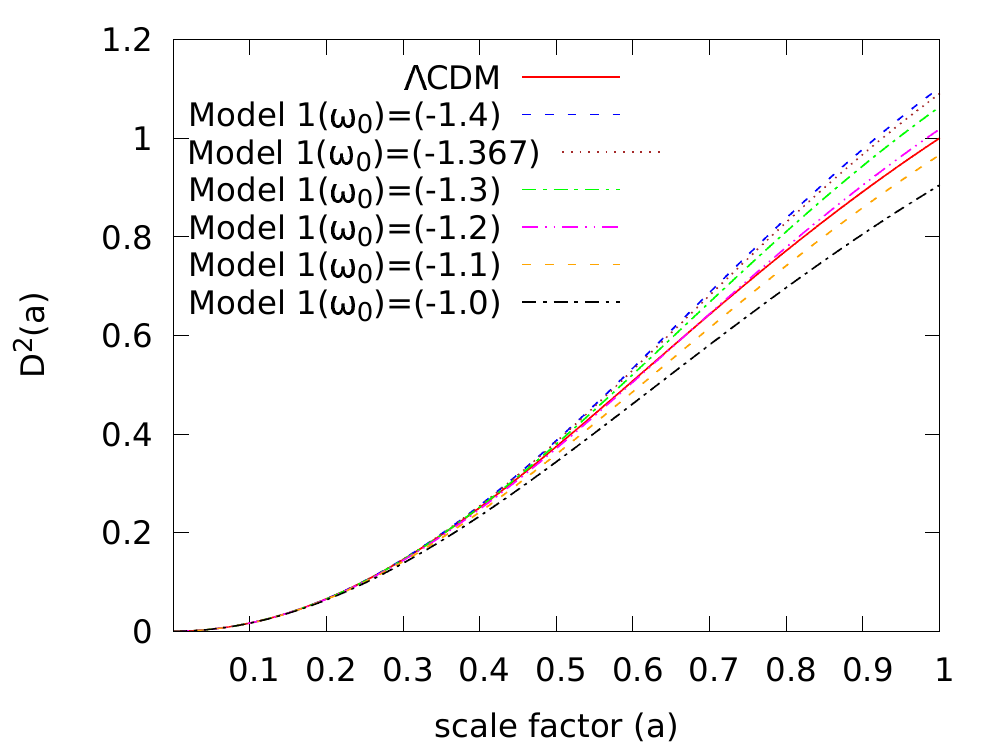}}}
  \hspace{0.5 cm}
  \resizebox{7 cm}{!}{\rotatebox{0}{\includegraphics{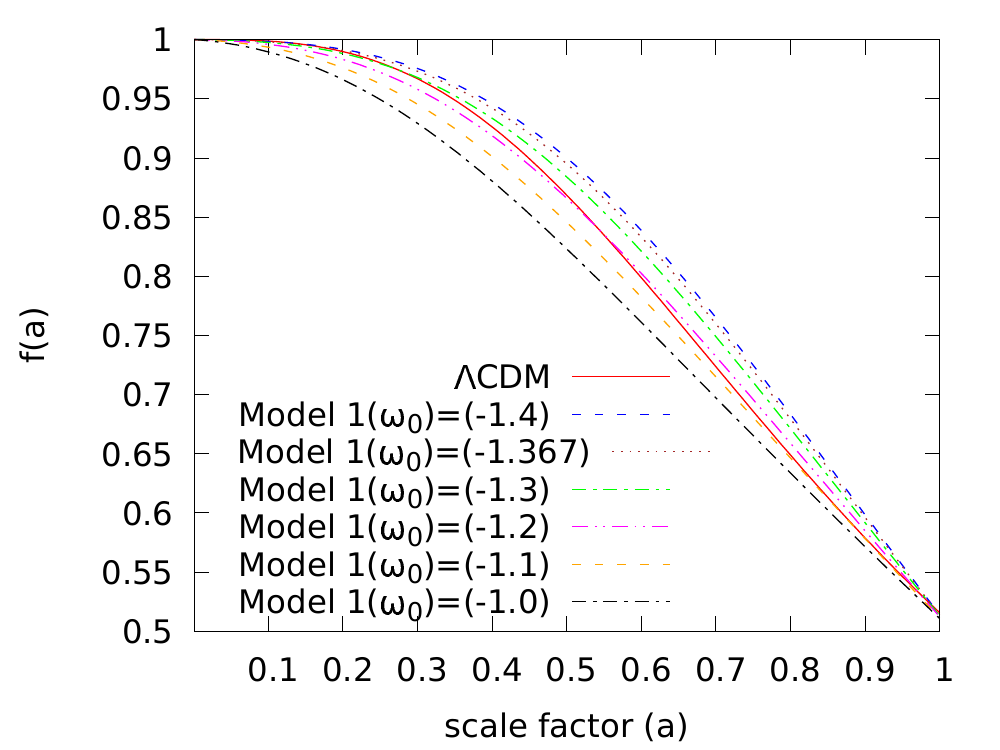}}}\\
  \vspace{-0.1 cm}   
   \resizebox{7 cm}{!}{\rotatebox{0}{\includegraphics{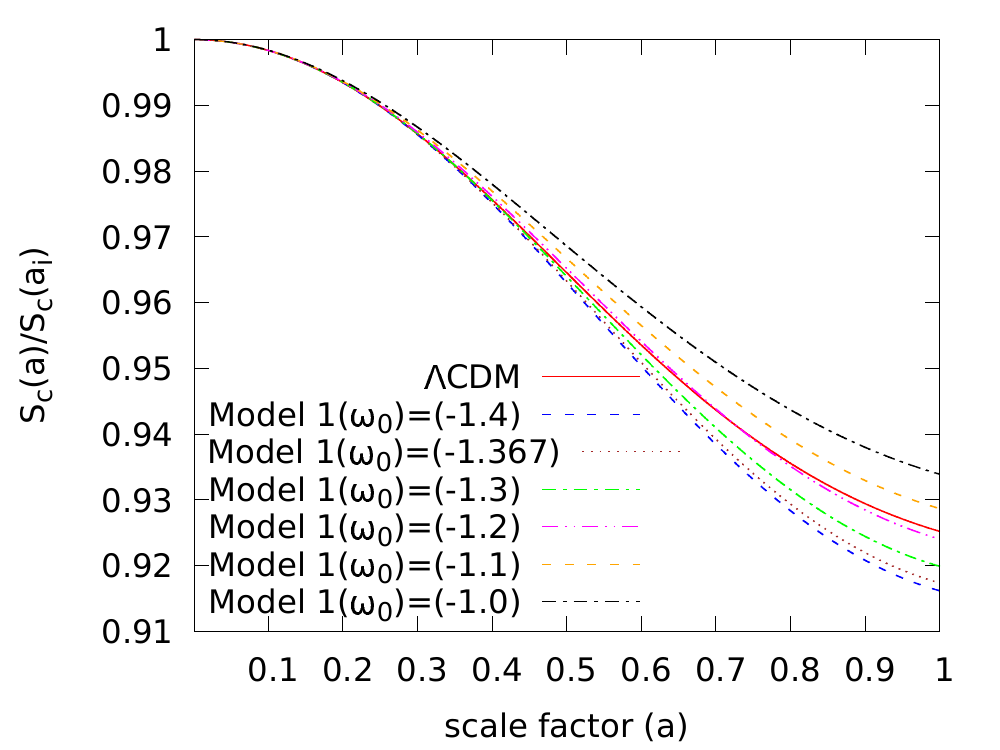}}}
  \hspace{0.5 cm}
  \resizebox{7 cm}{!}{\rotatebox{0}{\includegraphics{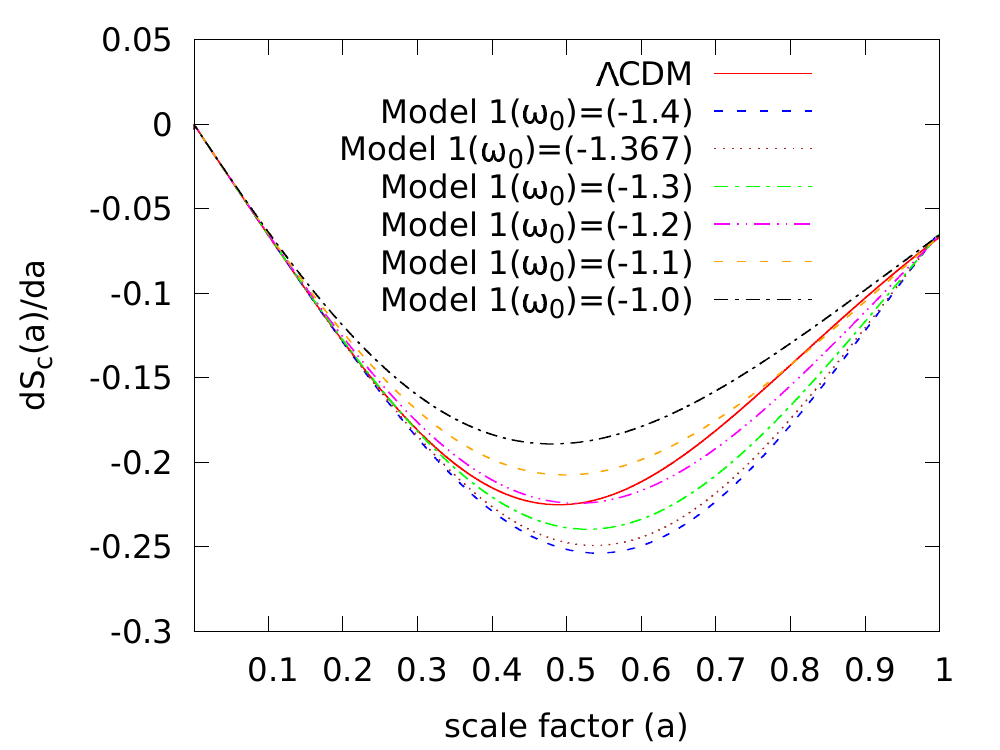}}}\\
   \vspace{-0.1 cm}
   \resizebox{7 cm}{!}{\rotatebox{0}{\includegraphics{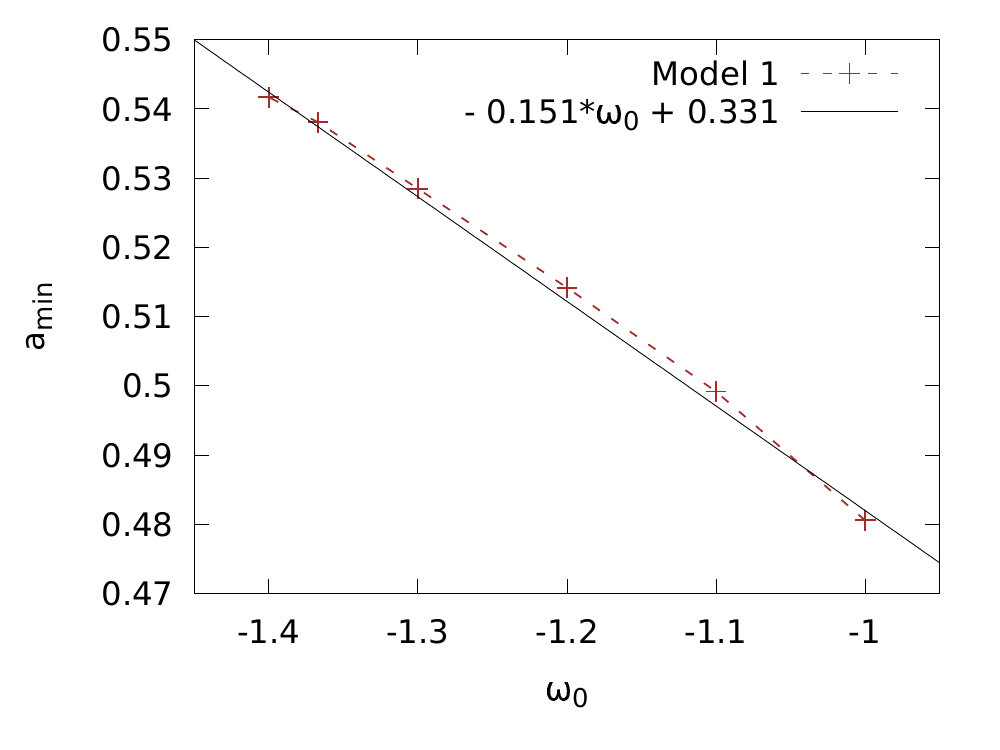}}}
   \hspace{0.5 cm}
   \resizebox{7 cm}{!}{\rotatebox{0}{\includegraphics{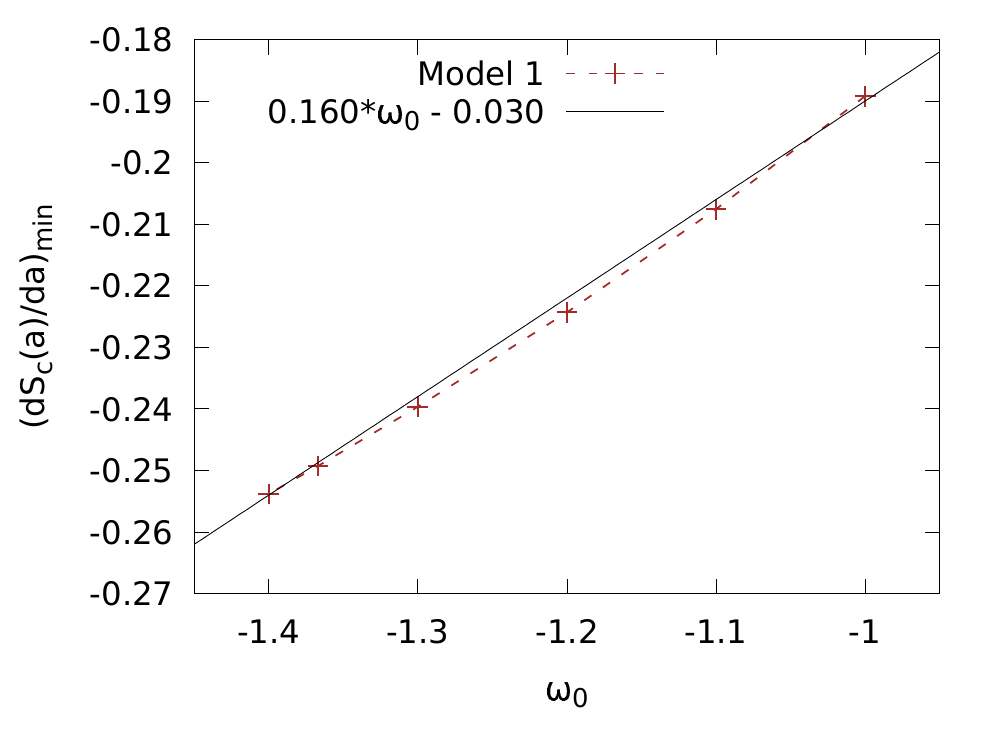}}}\\
   \caption{The top left and right panels respectively show $D^2(a)$
     and $f(a)$ for different values of $\omega_0$ in Model 1 along
     with the results from the $\Lambda$CDM model. The middle left
     panel shows the evolution of the configuration entropy
     $\frac{S_c(a)}{S_c(a_i)}$ with scale factor $a$ in Model 1. The
     middle right panel shows the entropy rate as a function of scale
     factor in Model 1. The different curves in these panels
     correspond to different values of $\omega_0$ in Model 1. The
     entropy rate in each of these models exhibits a minimum at a
     specific scale factor $a_{min}$. The value of $a_{min}$ and the
     magnitude of the entropy rate $[\frac{dS_c(a)}{da}]_{min}$ at
     $a_{min}$ depend on the value of the parameter $\omega_0$. We
     plot the dependence of $a_{min}$ on $\omega_0$ for Model 1 in the
     bottom left panel of this figure. The corresponding best fit line
     is also plotted together in the same panel. The dependence of
     $[\frac{dS_c(a)}{da}]_{min}$ on $\omega_{0}$ for Model 1 is shown
     in the bottom right panel. A best fit line describing this
     dependence is also shown together in the same panel.}
   \label{fig:one}
 \end{figure*}

 \begin{figure*}
   \resizebox{7 cm}{!}{\rotatebox{0}{\includegraphics{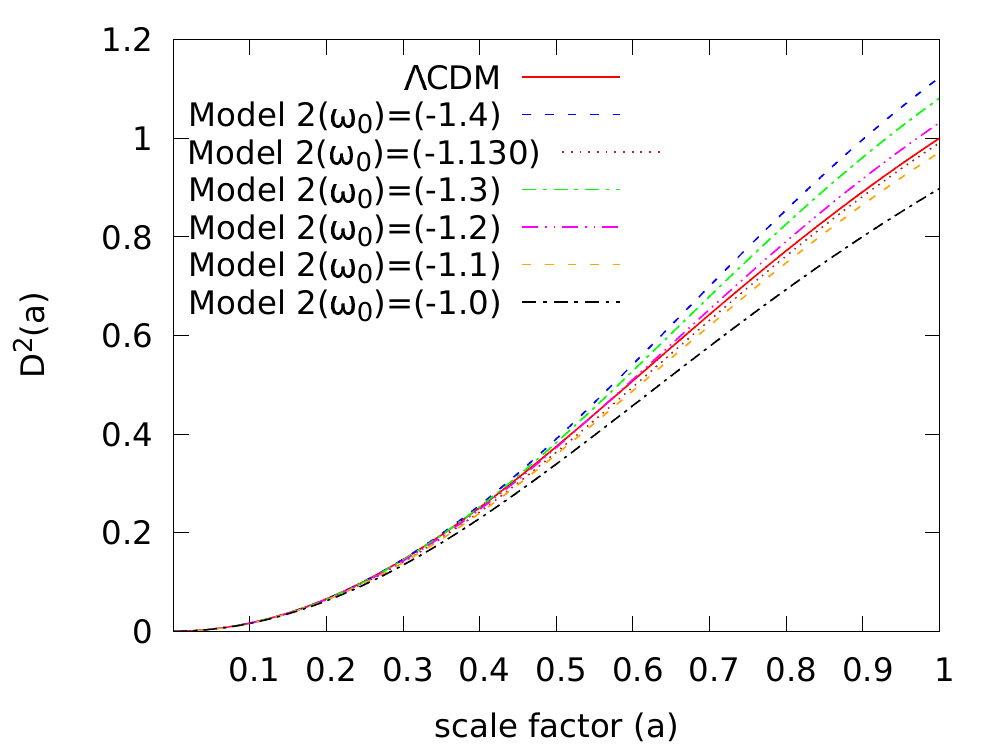}}}
  \hspace{0.5 cm}
  \resizebox{7 cm}{!}{\rotatebox{0}{\includegraphics{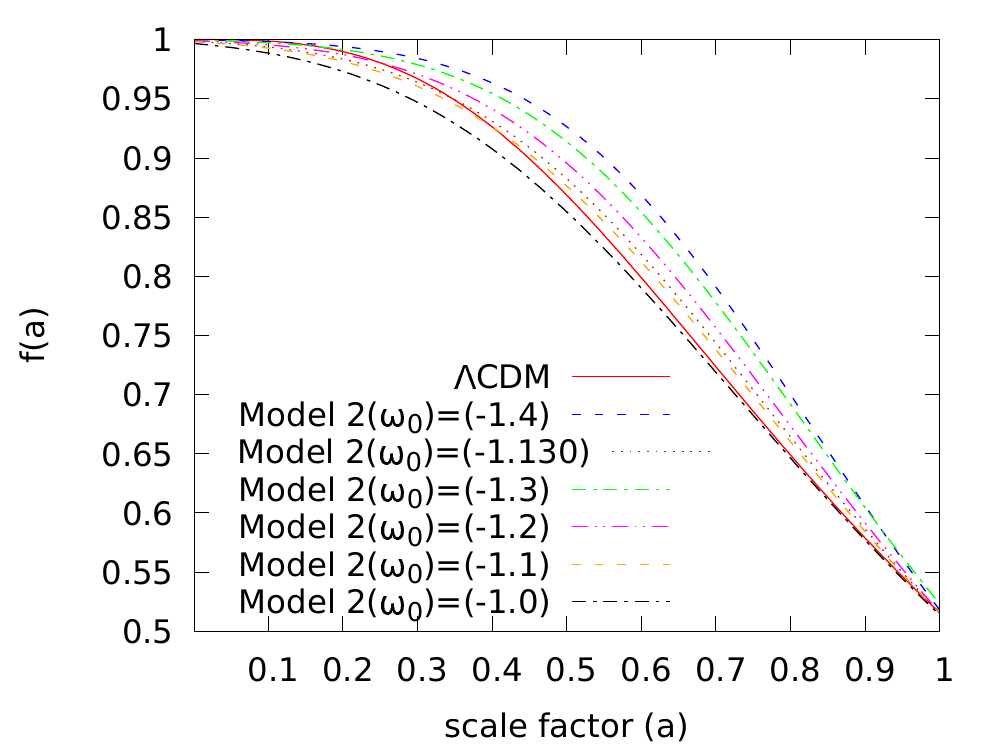}}}\\
  \vspace{-0.1 cm}  
   \resizebox{7 cm}{!}{\rotatebox{0}{\includegraphics{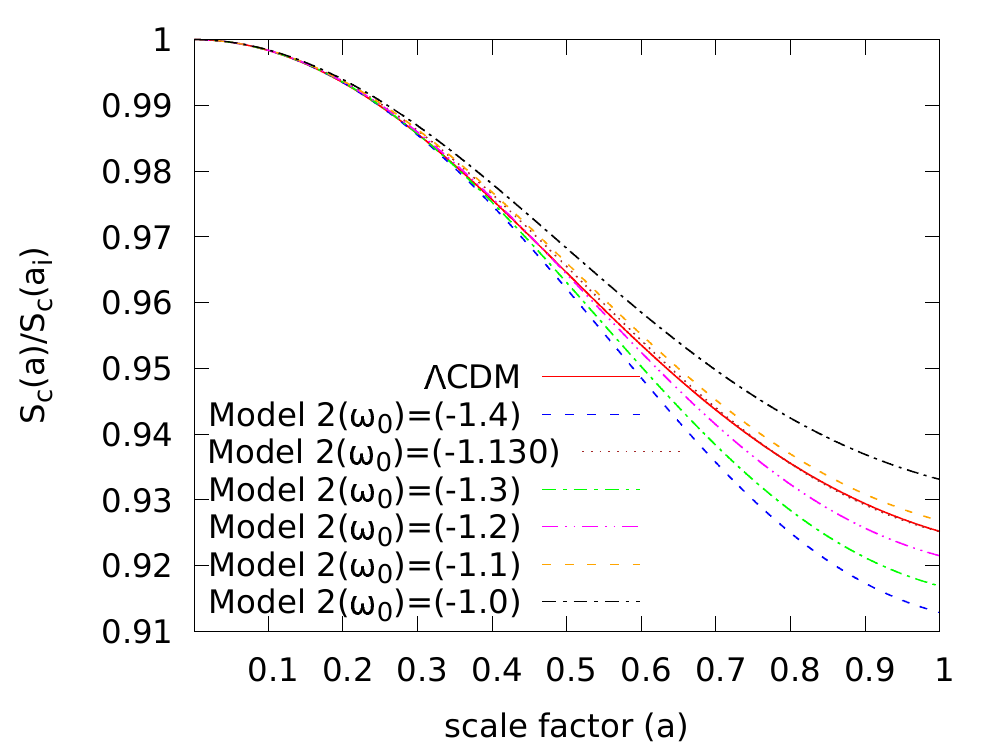}}}
  \hspace{0.5 cm}
  \resizebox{7 cm}{!}{\rotatebox{0}{\includegraphics{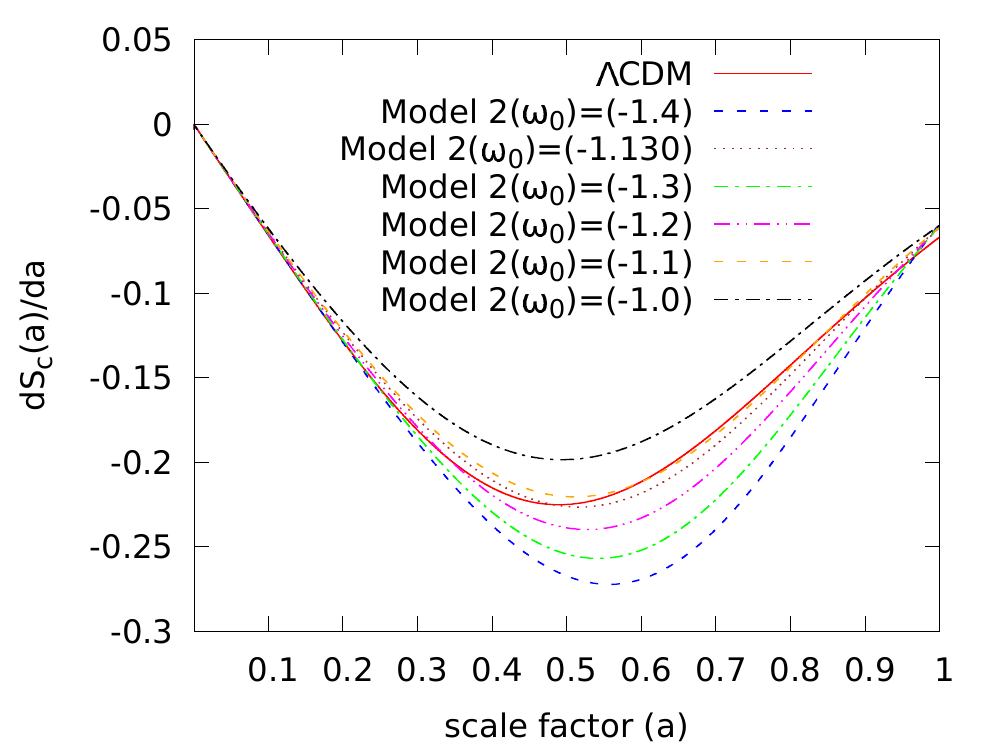}}}\\
   \vspace{-0.1 cm}
   \resizebox{7 cm}{!}{\rotatebox{0}{\includegraphics{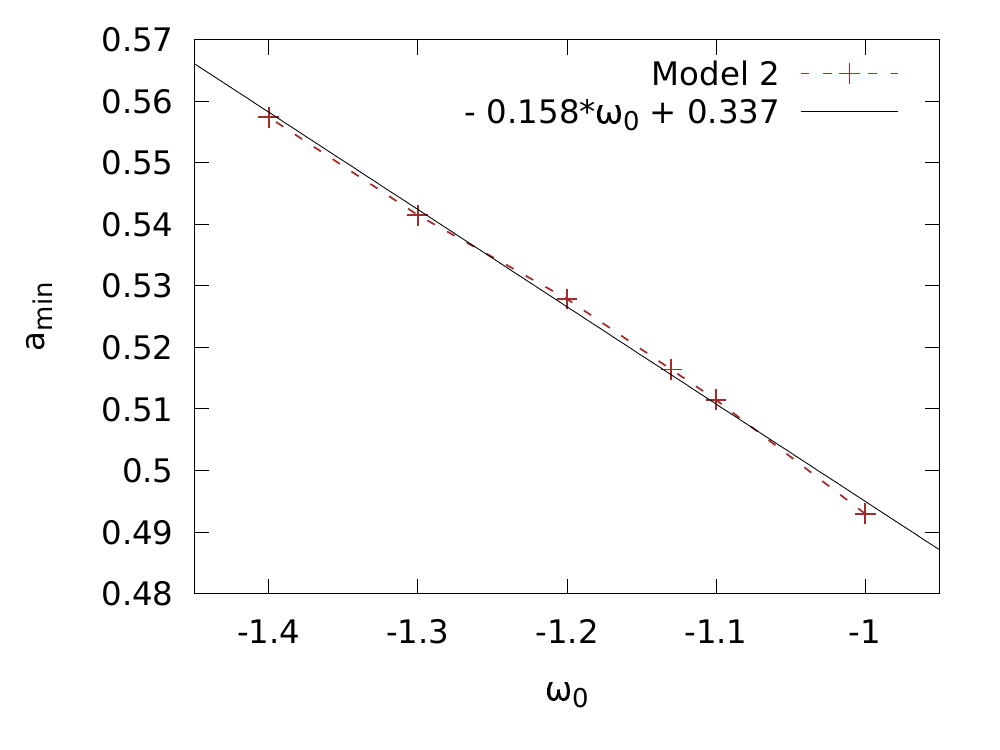}}}
   \hspace{0.5 cm}
   \resizebox{7 cm}{!}{\rotatebox{0}{\includegraphics{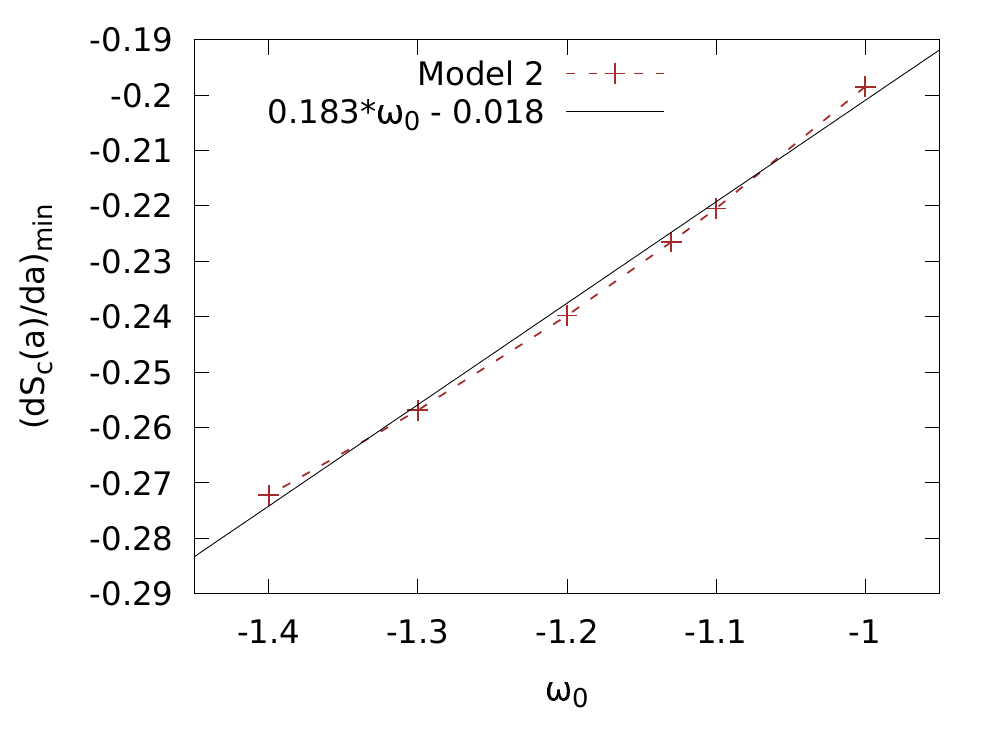}}}\\
   \caption{Same as \autoref{fig:one} but for Model 2.}
   \label{fig:two}
 \end{figure*}

  \begin{figure*}
   \resizebox{7 cm}{!}{\rotatebox{0}{\includegraphics{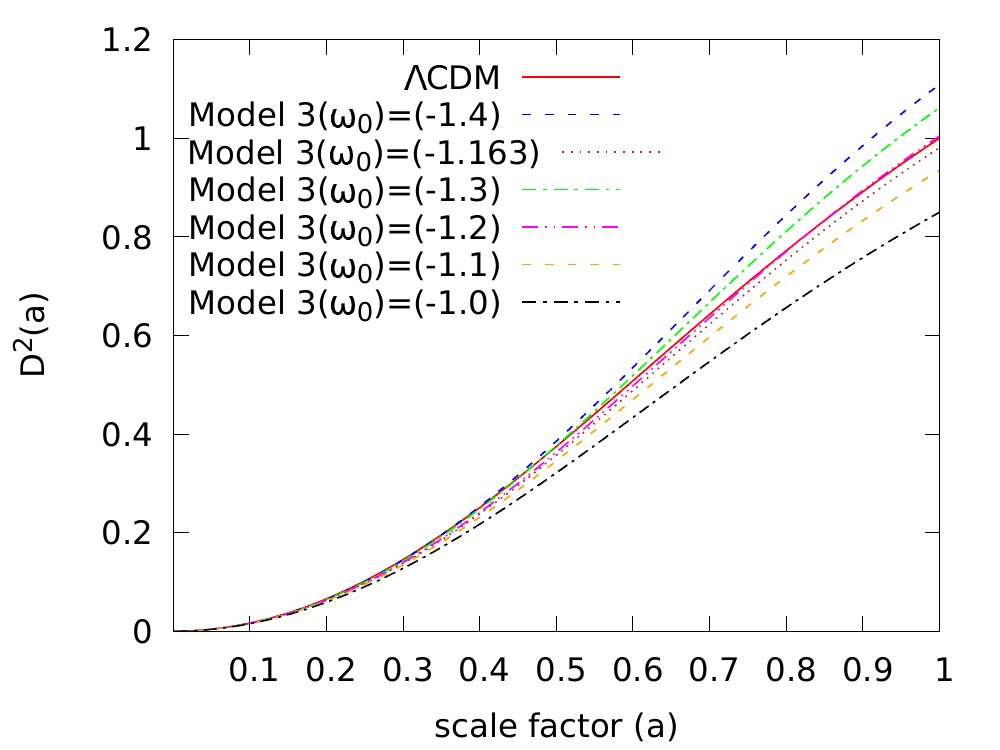}}}
  \hspace{0.5 cm}
  \resizebox{7 cm}{!}{\rotatebox{0}{\includegraphics{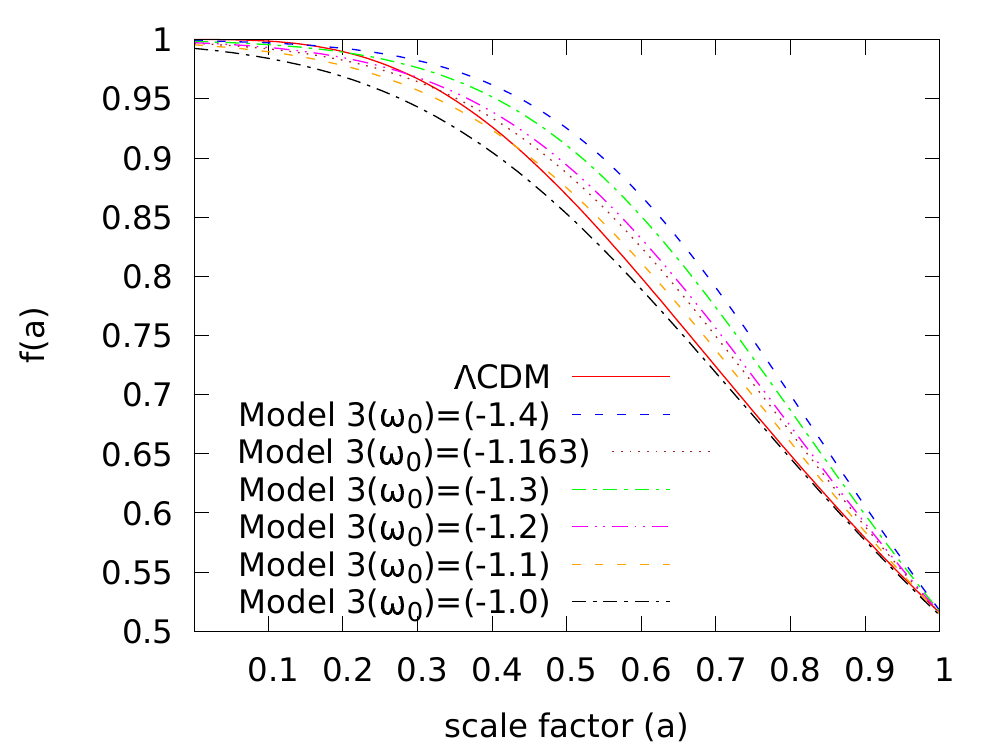}}}\\
  \vspace{-0.1 cm}  
   \resizebox{7 cm}{!}{\rotatebox{0}{\includegraphics{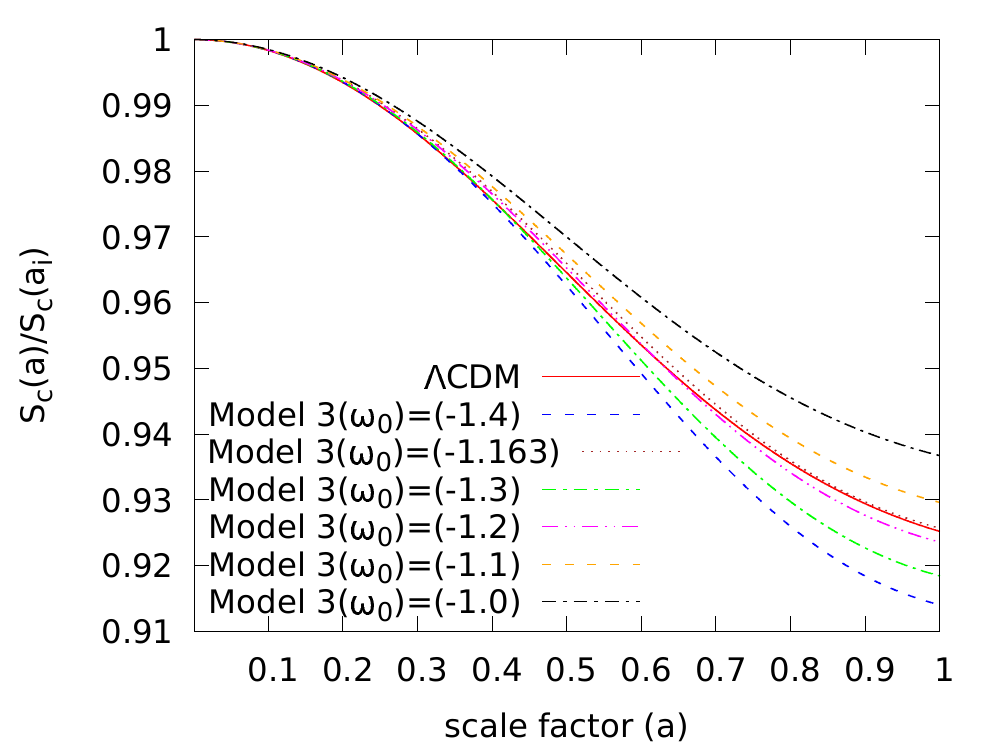}}}
  \hspace{0.5 cm}
  \resizebox{7 cm}{!}{\rotatebox{0}{\includegraphics{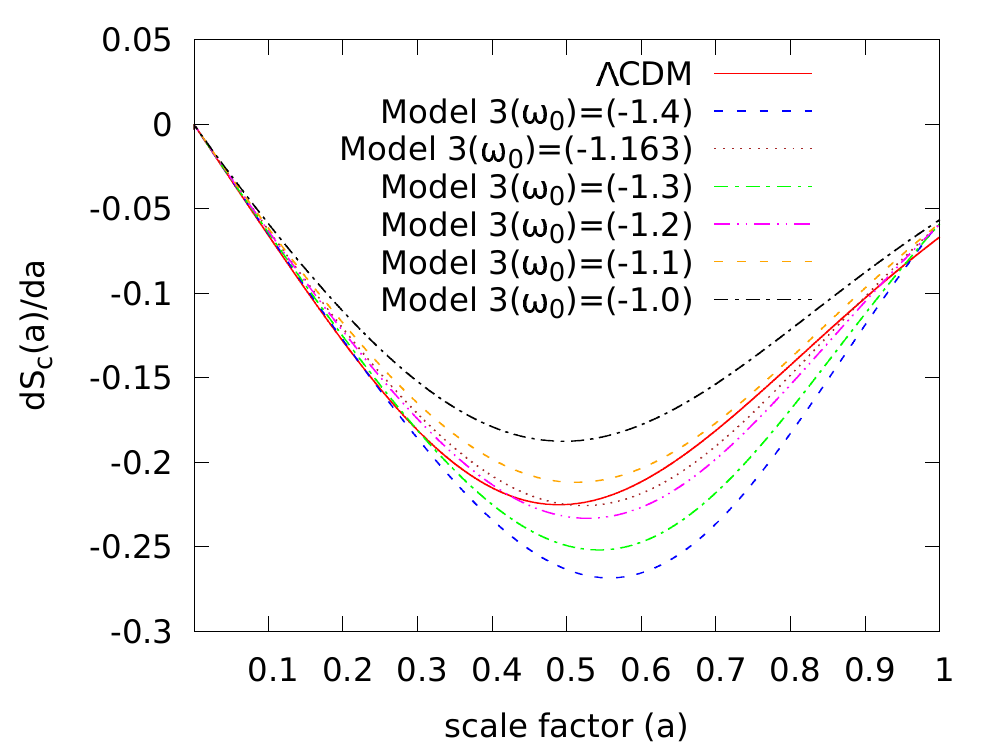}}}\\
   \vspace{-0.1 cm}
   \resizebox{7 cm}{!}{\rotatebox{0}{\includegraphics{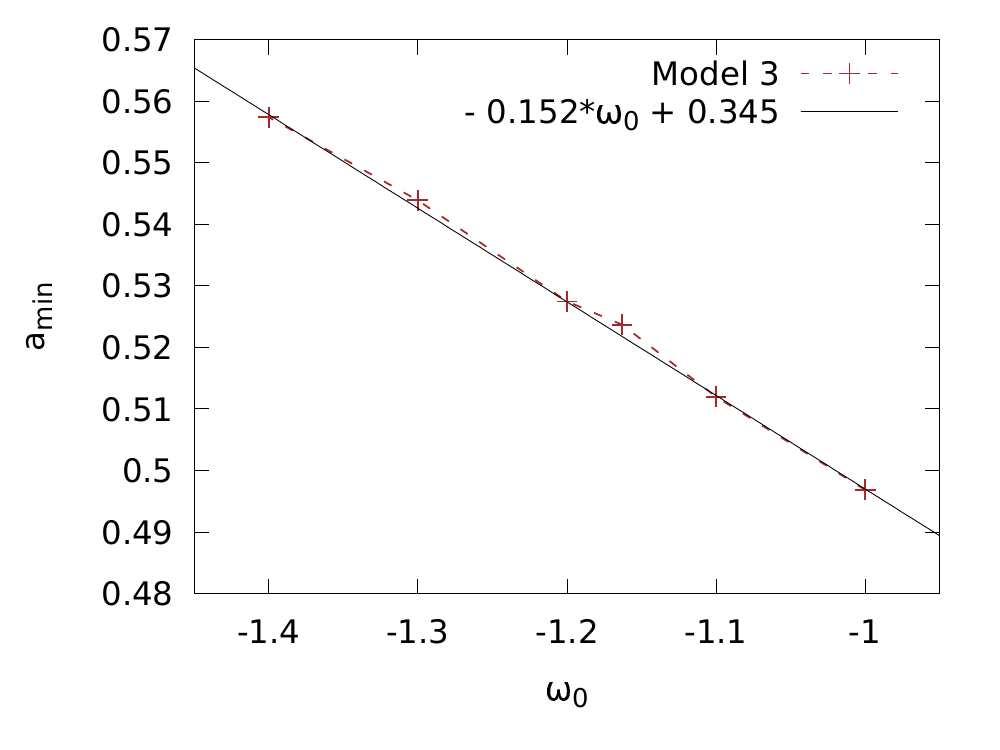}}}
   \hspace{0.5 cm}
   \resizebox{7 cm}{!}{\rotatebox{0}{\includegraphics{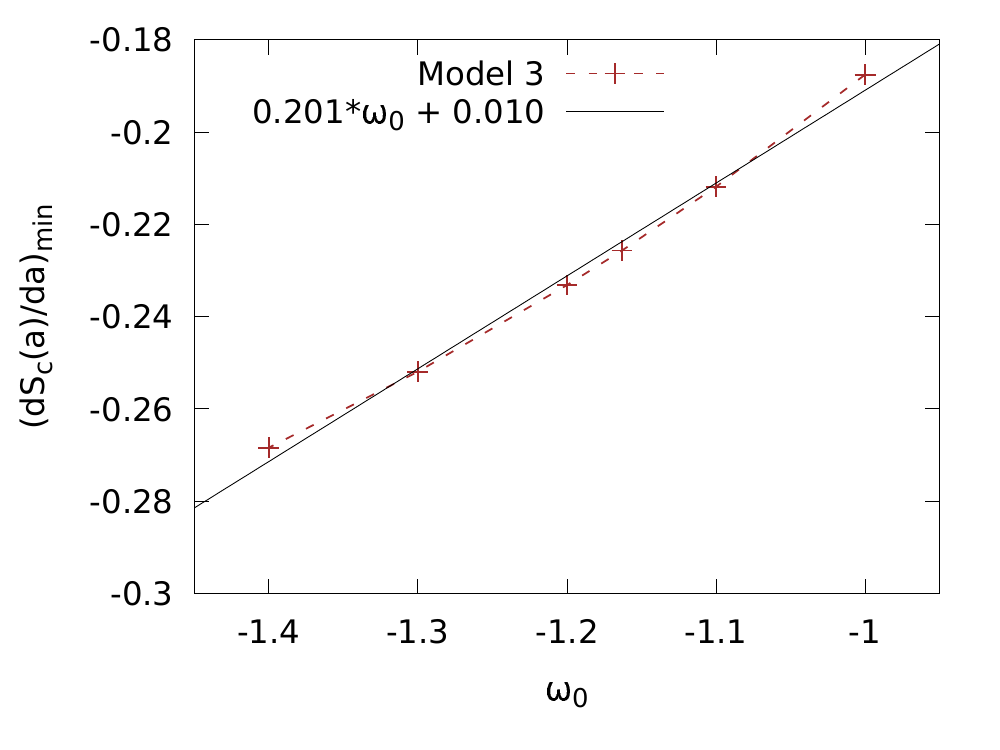}}}\\
   \caption{Same as \autoref{fig:one} but for Model 3.}
   \label{fig:three}
  \end{figure*}

   \begin{figure*}
  \resizebox{7 cm}{!}{\rotatebox{0}{\includegraphics{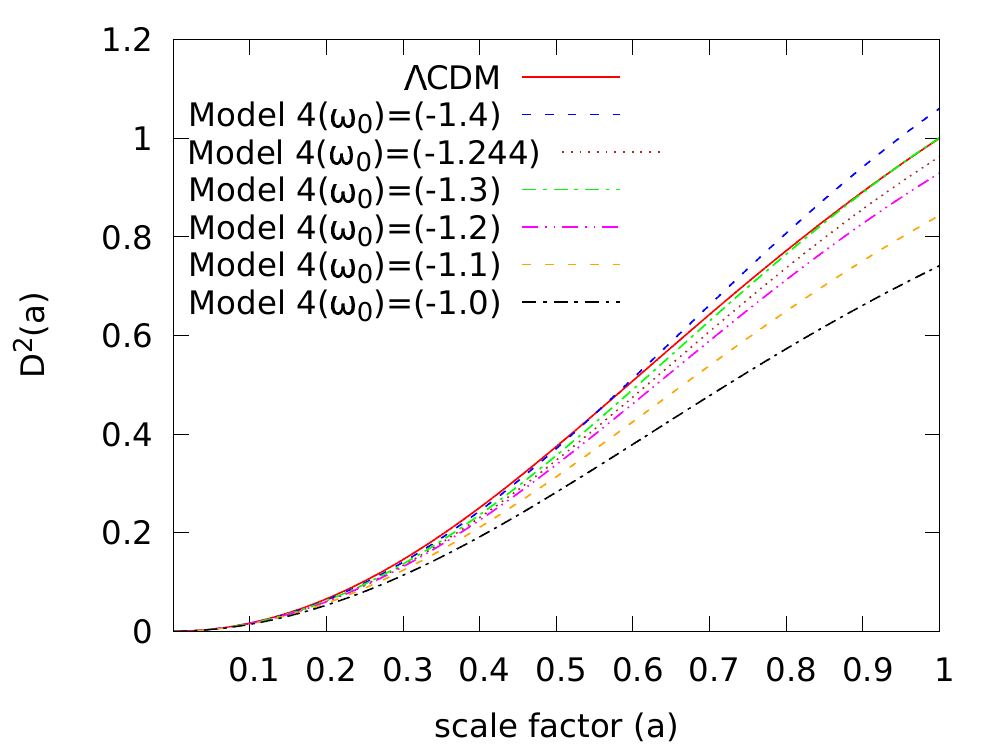}}}
  \hspace{0.5 cm}
  \resizebox{7 cm}{!}{\rotatebox{0}{\includegraphics{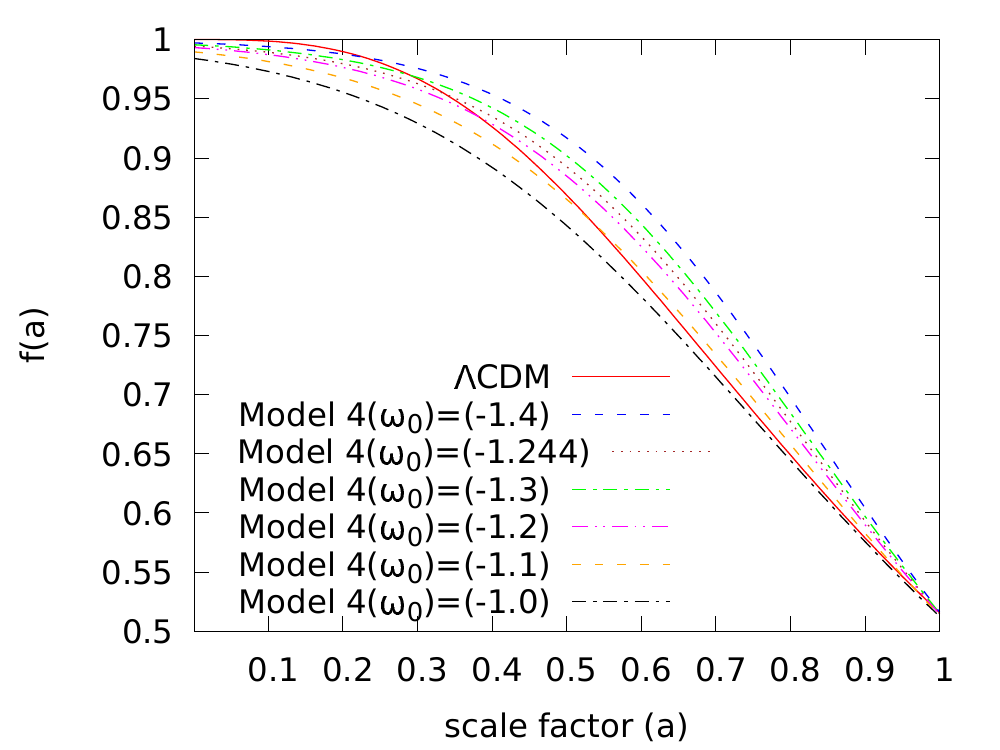}}}\\
  \vspace{-0.1 cm}  
   \resizebox{7 cm}{!}{\rotatebox{0}{\includegraphics{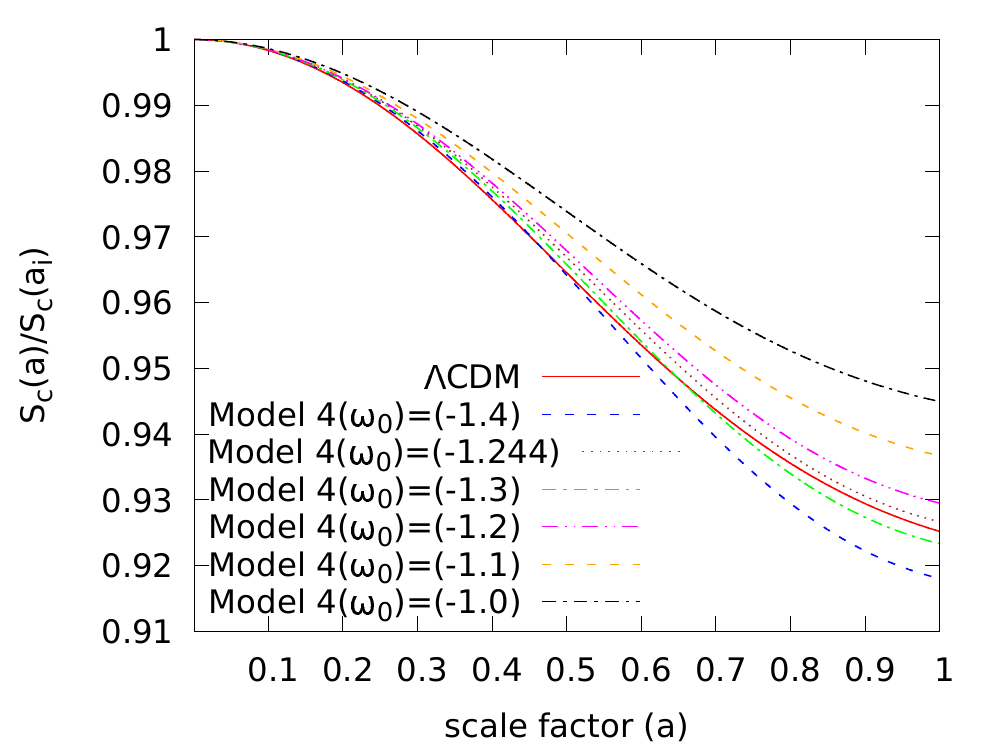}}}
  \hspace{0.5 cm}
  \resizebox{7 cm}{!}{\rotatebox{0}{\includegraphics{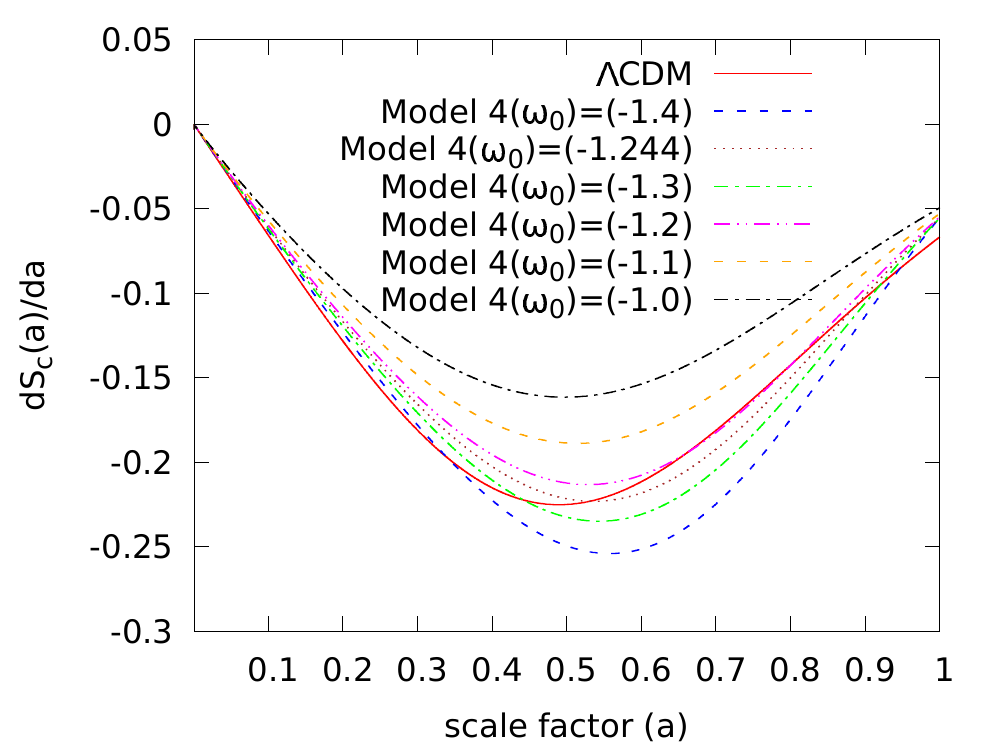}}}\\
   \vspace{-0.1 cm}
   \resizebox{7 cm}{!}{\rotatebox{0}{\includegraphics{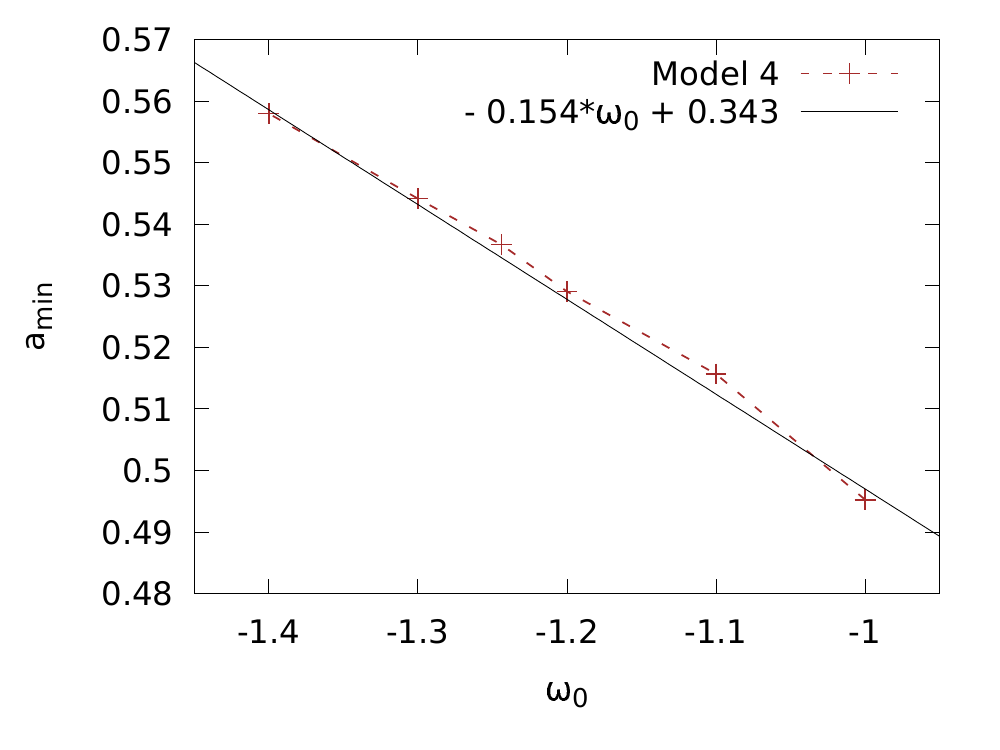}}}
   \hspace{0.5 cm}
   \resizebox{7 cm}{!}{\rotatebox{0}{\includegraphics{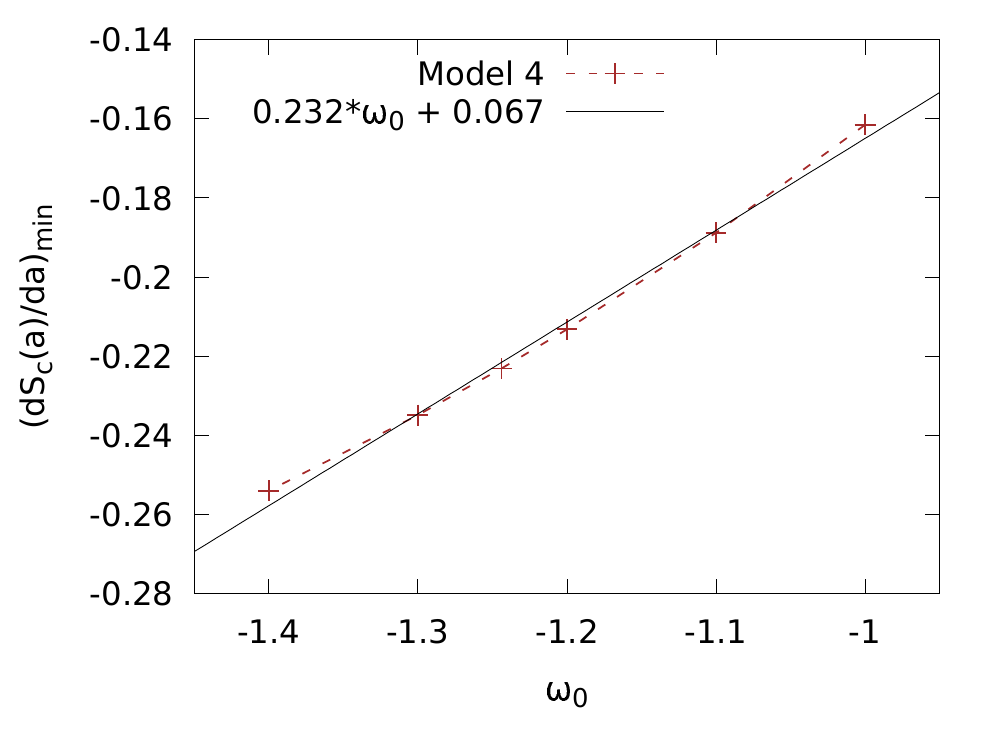}}}\\
   \caption{Same as \autoref{fig:one} but for Model 4.}
   \label{fig:four}
   \end{figure*}

    \begin{figure*}
   \resizebox{7 cm}{!}{\rotatebox{0}{\includegraphics{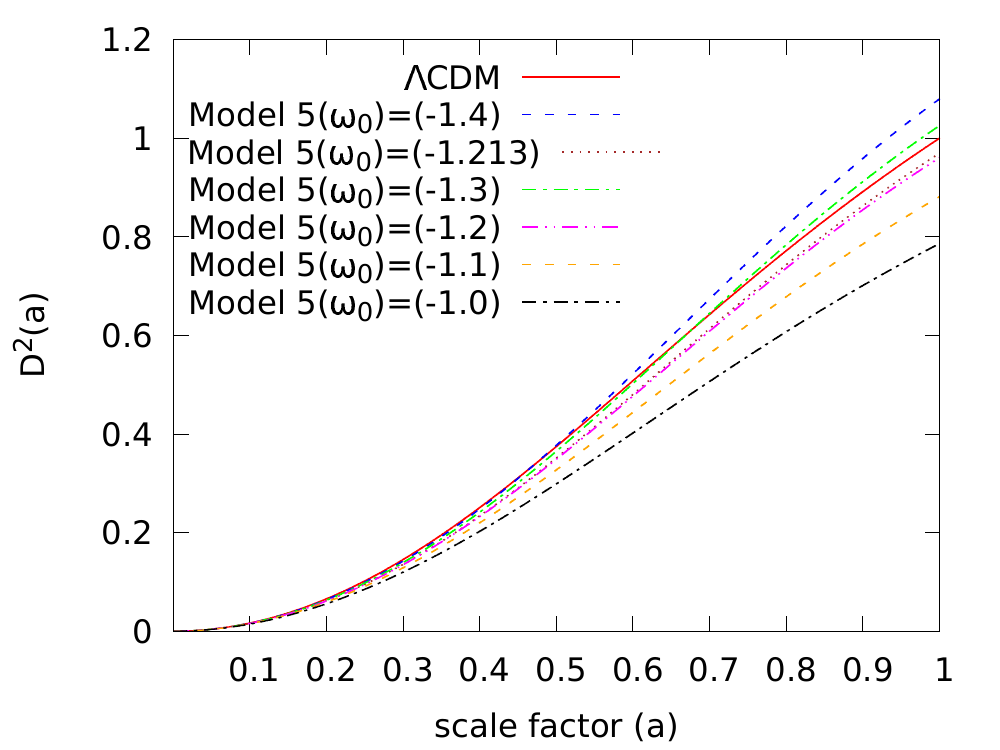}}}
   \hspace{0.5 cm}
   \resizebox{7 cm}{!}{\rotatebox{0}{\includegraphics{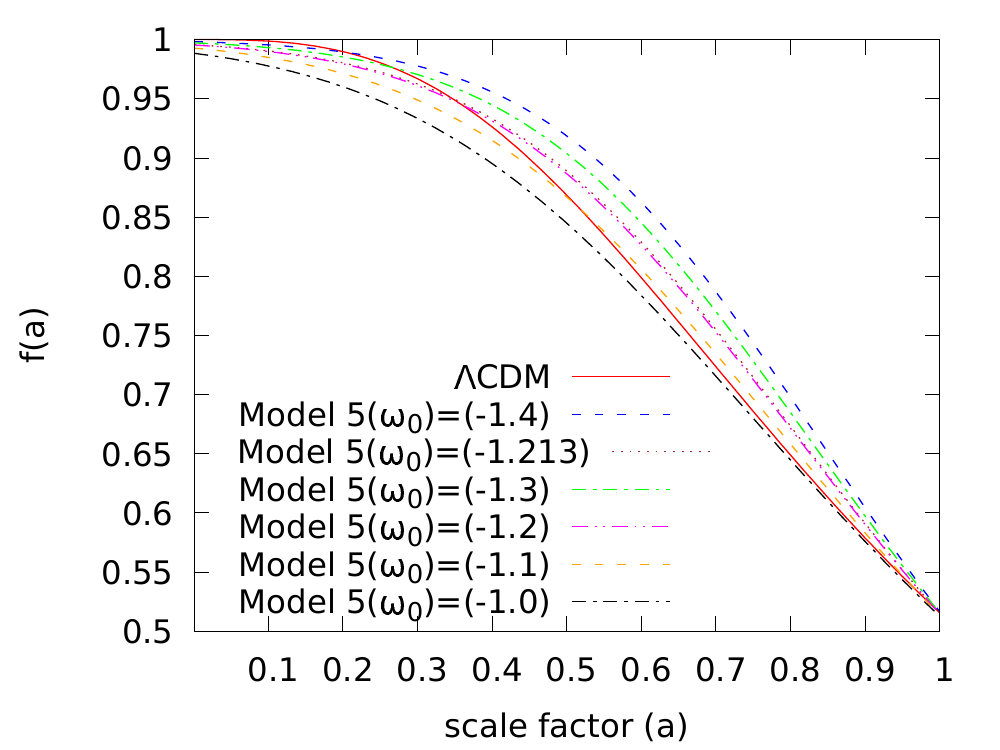}}}\\
   \vspace{-0.1 cm}  
   \resizebox{7 cm}{!}{\rotatebox{0}{\includegraphics{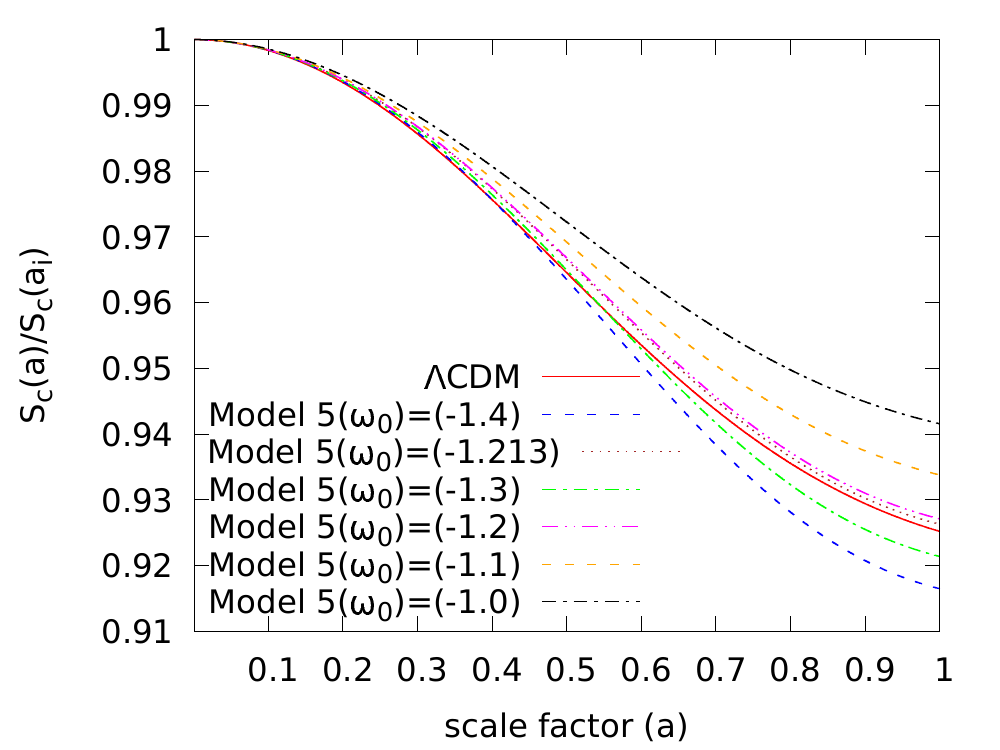}}}
  \hspace{0.5 cm}
  \resizebox{7 cm}{!}{\rotatebox{0}{\includegraphics{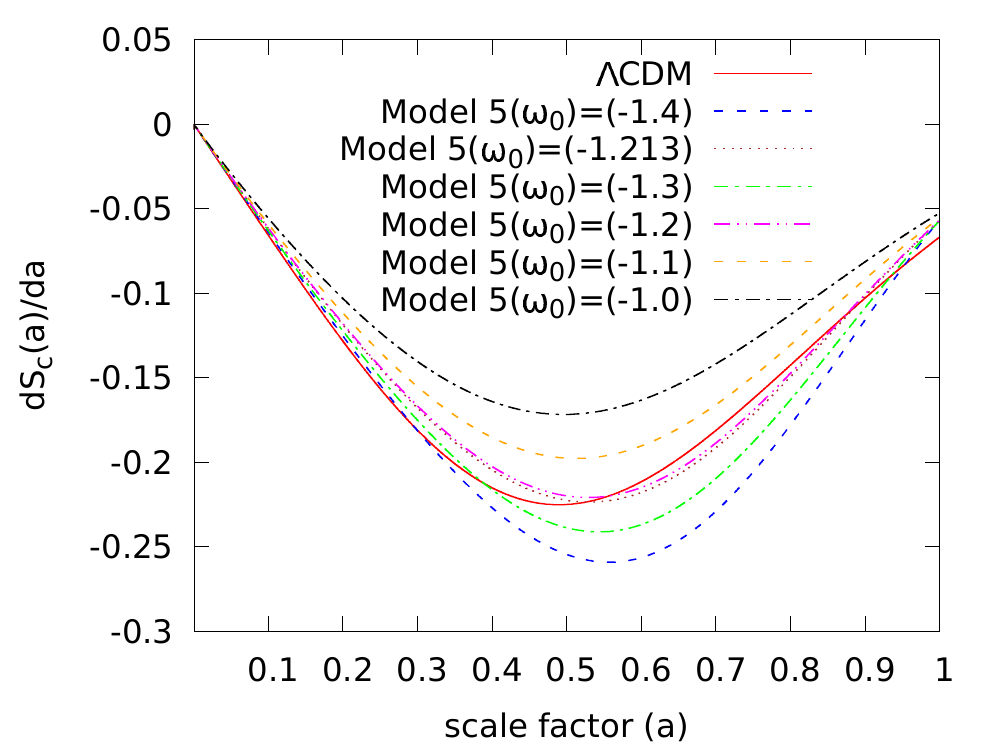}}}\\
   \vspace{-0.1 cm}
   \resizebox{7 cm}{!}{\rotatebox{0}{\includegraphics{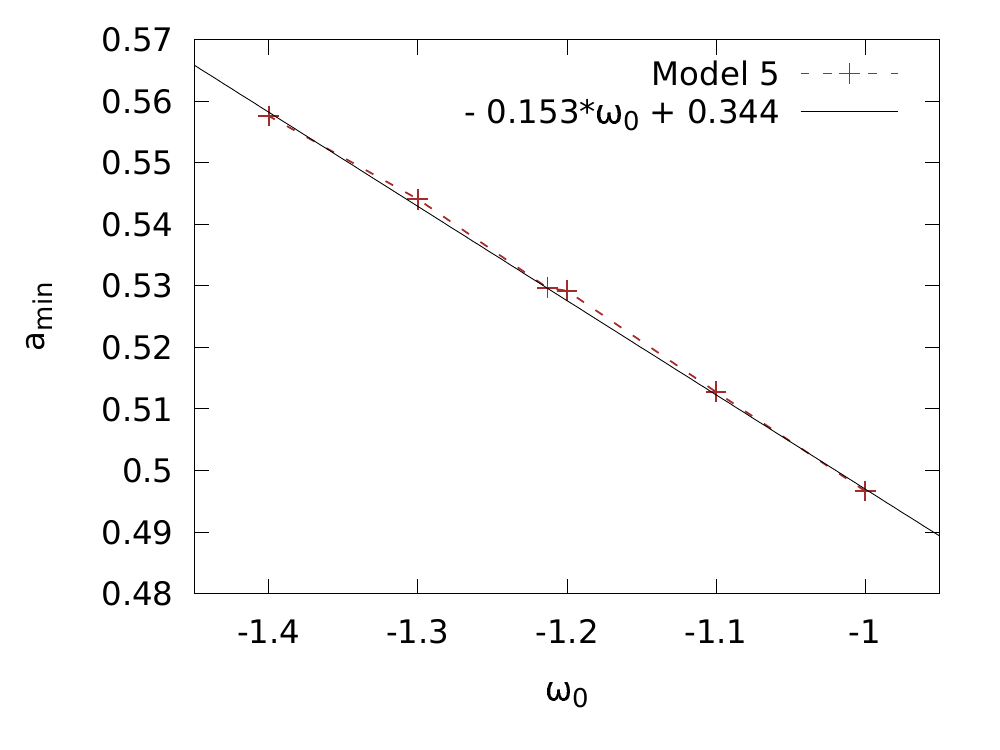}}}
   \hspace{0.5 cm}
   \resizebox{7 cm}{!}{\rotatebox{0}{\includegraphics{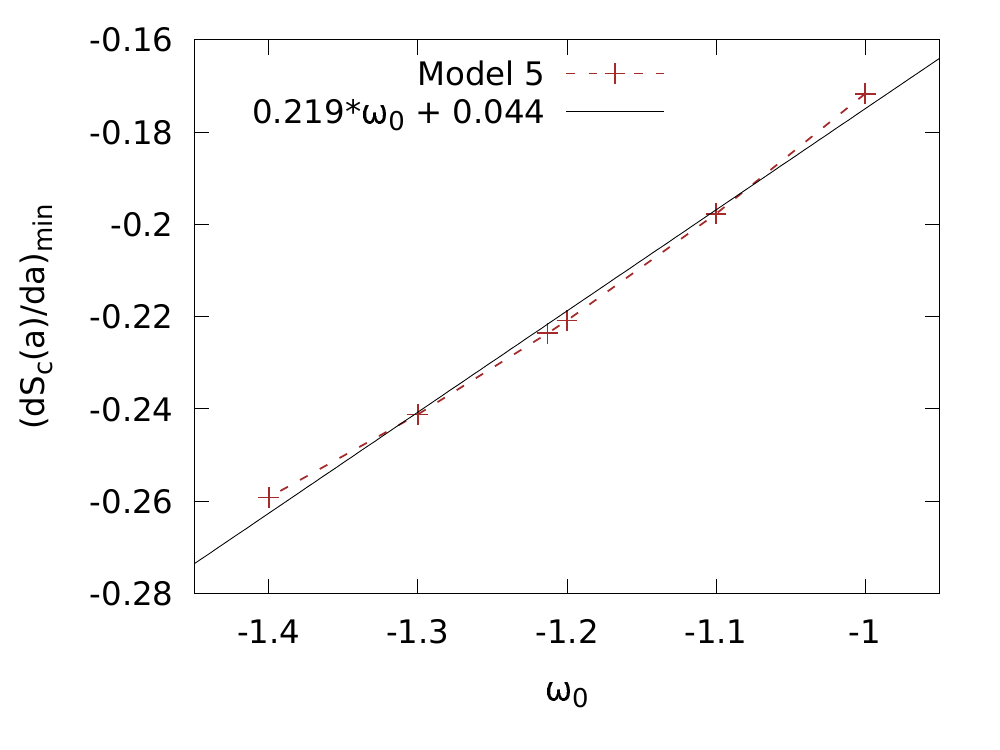}}}\\
   \caption{Same as \autoref{fig:one} but for Model 5.}
   \label{fig:five}
 \end{figure*}

\section{Theory}
\subsection{Evolution of configuration entropy}

We consider a large comoving volume $V$ and divide it into a number of
identical sub-volumes $dV$. If at any instant $t$, the density $\rho
(\vec{x}, t)$ inside each of these sub-volumes are known then the
configuration entropy of the mass distribution in the volume $V$ can
be written as \citep{pandey1},
 \begin{eqnarray}
	  S_c(t) = - \int \rho (\vec {x}, t) \log \rho (\vec {x}, t)\, dV.
	  \label{eq:one}
  \end{eqnarray}
This definition is motivated by the idea of the information entropy
which was originally proposed by \citet{shannon48}.

 Treating the mass distribution as an ideal fluid, the continuity
 equation in an expanding Universe is given by,
  \begin{eqnarray}
	  \frac{\partial \rho}{\partial t} + 3 \frac{\dot a}{a} \rho +
          \frac{1}{a} \nabla \cdot (\rho \vec {v})=0.
	  \label{eq:two}
  \end{eqnarray}
  Here $a$ is the scale factor and $\vec {v}$ is the peculiar velocity
  of the fluid elements.
  
  The evolution of the configuration entropy \citep {pandey1} in
  volume $V$ can be obtained from \autoref{eq:two} as,
  \begin{eqnarray}
	  \frac{dS_c(t)}{dt} + 3 \frac{\dot a}{a} S_c(t) - \frac{1}{a}
          \int \rho (3 \dot a + \nabla \cdot \vec {v})\, dV = 0.
	  \label{eq:three}
  \end{eqnarray}
  The \autoref {eq:three} can be also written as,
  \begin{eqnarray}
    \frac{dS_c(a)}{da}\dot a + 3 \frac{\dot a}{a}S_c(a) - F(a) = 0,
	  \label{eq:four}
  \end{eqnarray}
  where, 
  \begin{eqnarray}
	  F(a) = 3MH(a) + \frac{1}{a} \int \rho (\vec {x}, a) \nabla
          \cdot \vec {v}\, dV.
	  \label{eq:five}
  \end{eqnarray}
  Here $H(a)$ is the Hubble parameter and $M = \int \rho (\vec {x},
  a)\, dV = \int \bar \rho (1 + \delta (\vec {x}, a))\, dV$ is the
  total mass inside the comoving volume $V$.  $\bar \rho$ is the
  average density of matter within the comoving volume $V$ and
  $\delta(\vec {x}, a) = \frac{\rho (\vec {x}, a) - \bar \rho}{\bar \rho}$ is
  the density contrast at comoving coordinate $\vec {x}$ at time $t$.
  One can simplify \autoref{eq:four} further using the linear
  perturbation theory and get,
  \begin{eqnarray}
    \frac{dS_c(a)}{da} + \frac{3}{a} (S_c(a) - M) +
    \bar \rho f(a) \frac{D^2(a)}{a} \int \delta^2 (\vec {x})\, dV = 0.
	  \label{eq:six}
  \end{eqnarray}
  Here $D(a)$ is the growing mode of density perturbations and $f(a) =
  \frac {d \ln D}{d \ln a}$ is the dimensionless linear growth rate.

  We need to solve \autoref{eq:six} to find the evolution of entropy
  as a function of scale factor. We first require $D(a)$ and $f(a)$ to
  solve \autoref{eq:six}. These are cosmology dependent quantities which
  have to be evaluated separately for each specific model under
  consideration. For simplicity, we set the time independent quantities
  equal to $1$ in \autoref{eq:six} and solve the equation using fourth
  order Runge-Kutta method. 

   The entropy evolution is jointly determined by the the second and
   third term of \autoref{eq:six}. The second term is decided by the
   initial condition and the third term is primarily determined by
   growth rate of structure formation. Since at very early times
   growth rate is negligible, entropy evolution in this period is
   almost completely determined by the initial condition. An
   analytical solution of \autoref{eq:six} ignoring the third term is
   given by,
  \begin{eqnarray}
	  \frac{S_c(a)}{S_c(a_i)} = \frac{M}{S_c(a_i)} + \left[1 - \frac{M}{S_c(a_i)}\right]\left(\frac{a_i}{a}\right)^3.
	  \label{eq:eight}
  \end{eqnarray}
  Here $a_i$ is the initial scale factor and $S_c(a_i)$ is the
  entropy at the initial scale factor. We choose $a_i=10^{-3}$
  throughout the analysis. The \autoref{eq:eight} suggests a sudden
  growth in $\frac{S_c(a)}{S_c(a_i)}$ near $a_i$ for $S_c(a_i) <
  M$. Similarly a sudden drop in the value of
  $\frac{S_c(a)}{S_c(a_i)}$ is expected near $a_i$ for $S_c(a_i) > M$.
  These transients have nothing to do with the cosmological model
  concerned. The choice of the initial condition is arbitrary. We set
  $S_c(a_i) = M$ throughout the present analysis to ignore the initial
  transients caused by the initial conditions.

  The third term in \autoref{eq:six} becomes important only after the
  significant growth of structures. The goal of the present analysis
  is to explore the possibility of constraining the dark energy EoS
  parameters using the evolution of configuration entropy. The dark
  energy equation of state influences the growth rate of structures
  and hence the cosmology dependent third term in \autoref{eq:six}
  will be of our primary interest. The time derivative of the
  configuration entropy can be obtained by simply using
  \autoref{eq:six} or by numerical differentiation of the solution of
  \autoref{eq:six}.

\begin{figure*}

   \resizebox{7 cm}{!}{\rotatebox{0}{\includegraphics{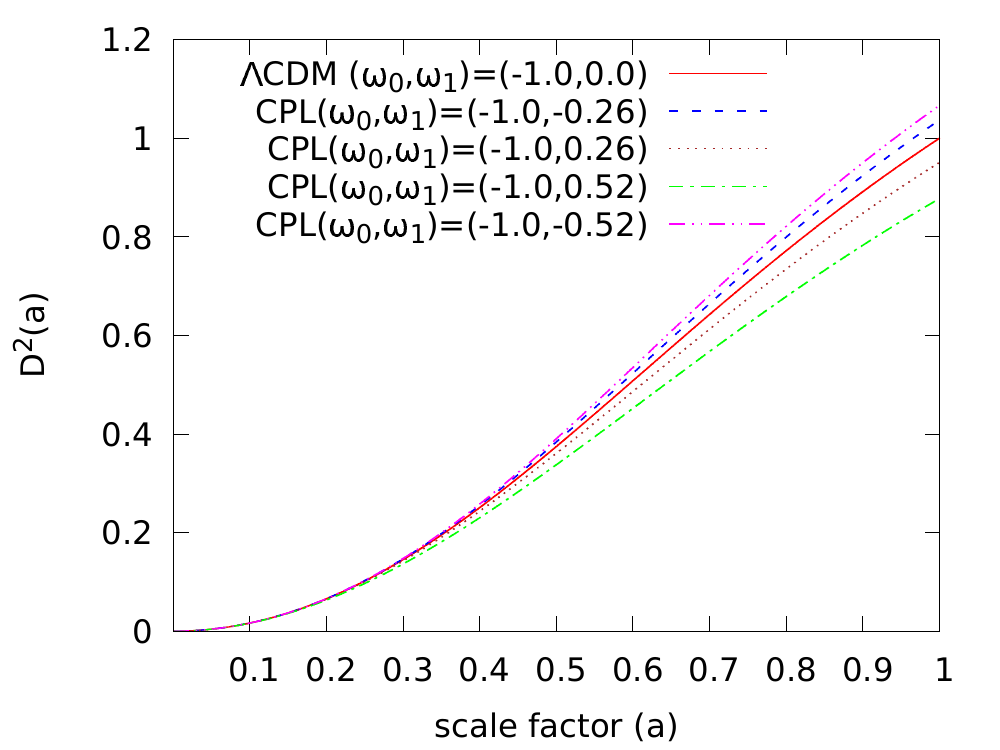}}}
  \hspace{0.5 cm}
  \resizebox{7 cm}{!}{\rotatebox{0}{\includegraphics{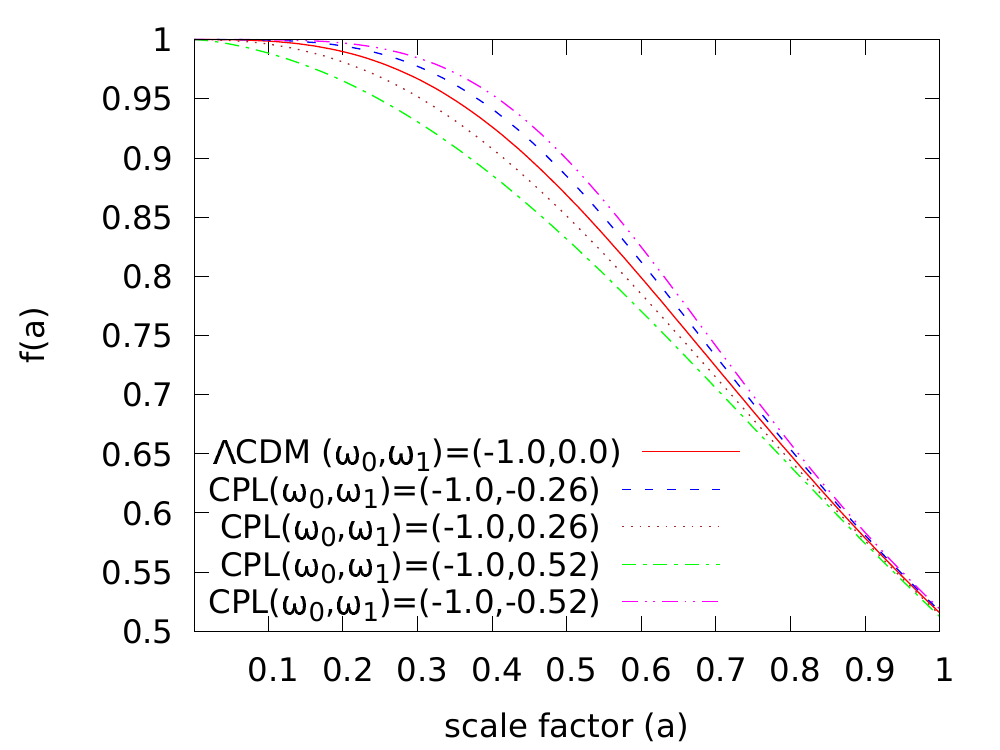}}}\\
   \vspace{-0.1 cm}
   \resizebox{7 cm}{!}{\rotatebox{0}{\includegraphics{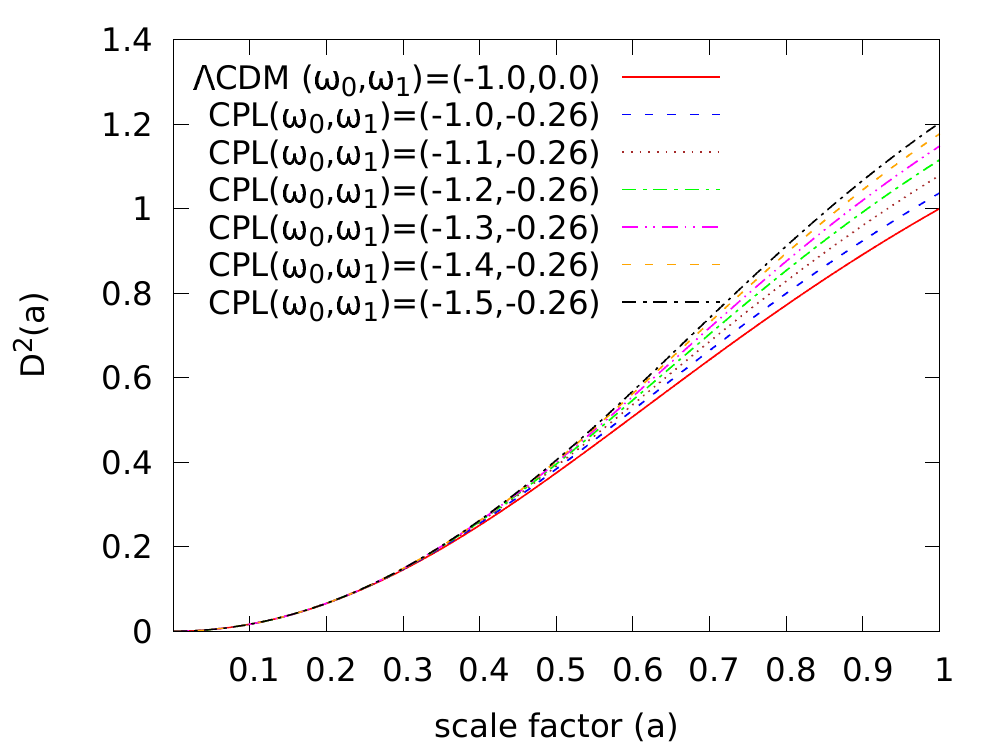}}}
   \hspace{0.5 cm}
   \resizebox{7 cm}{!}{\rotatebox{0}{\includegraphics{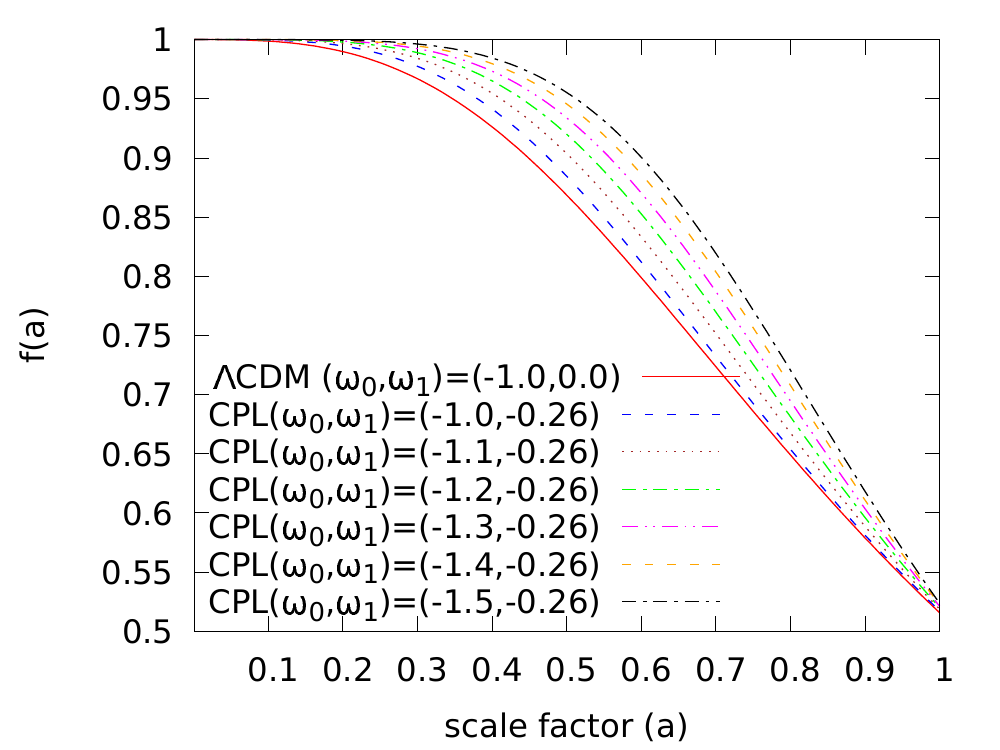}}}\\
   \caption{The top left panel shows $D^2(a)$ for CPL parametrization
     with a fixed value of $\omega_0$ but different values of
     $\omega_1$. The top right panel shows $f(a)$ for the same models
     shown in the top left panel. The bottom left panel and the bottom
     right panel respectively show $D^2(a)$ and $f(a)$ for CPL
     parametrization with a fixed value of $\omega_1$ but different
     values of $\omega_0$.}
   \label{fig:cpl_fd}
  \end{figure*}

\begin{figure*}

   \resizebox{7 cm}{!}{\rotatebox{0}{\includegraphics{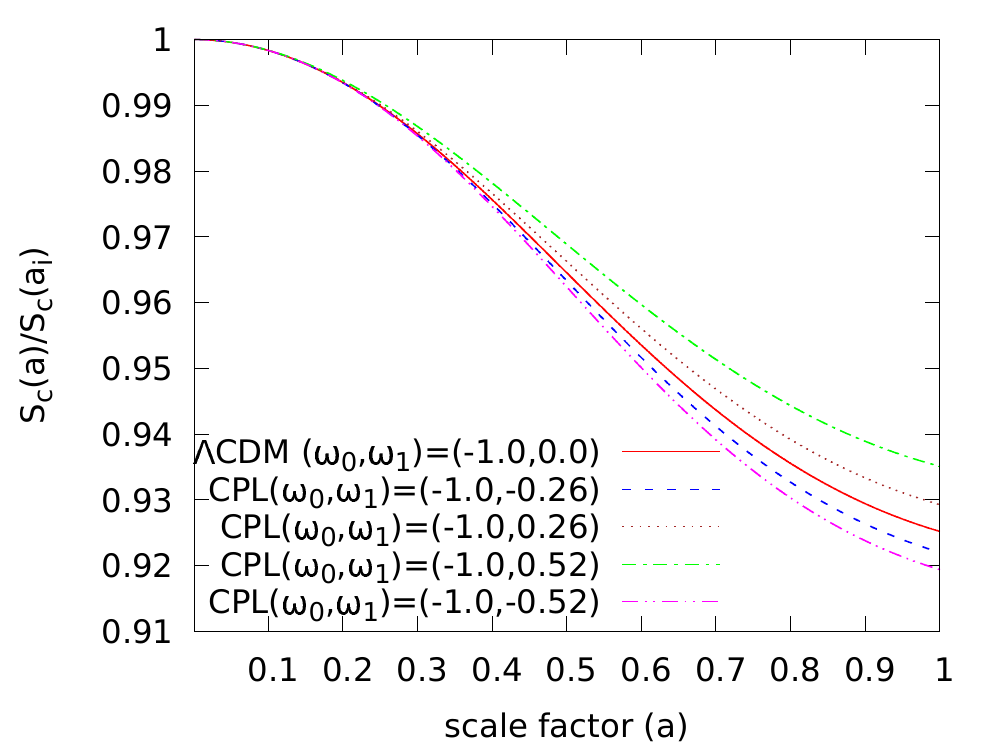}}}
  \hspace{0.5 cm}
  \resizebox{7 cm}{!}{\rotatebox{0}{\includegraphics{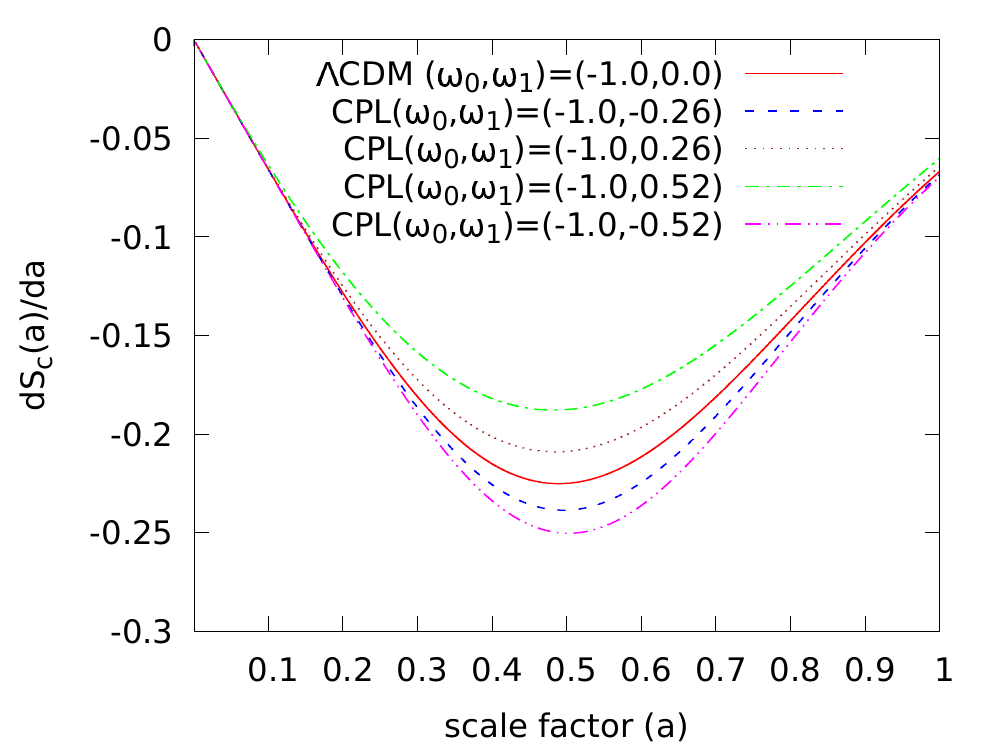}}}\\
   \vspace{-0.1 cm}
   \resizebox{7 cm}{!}{\rotatebox{0}{\includegraphics{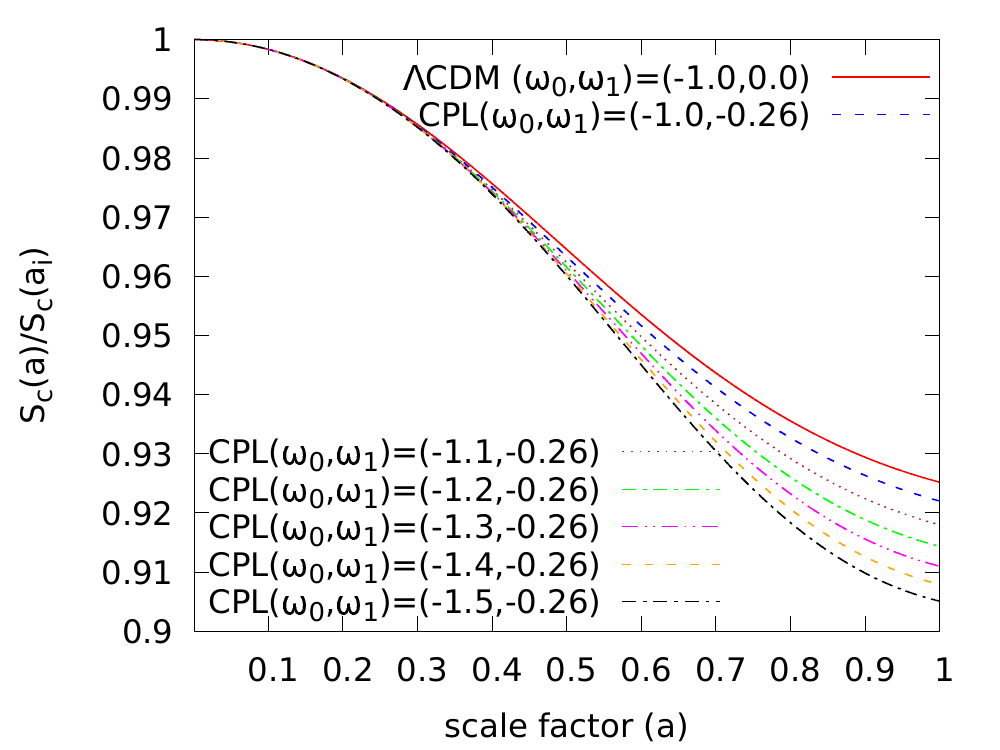}}}
   \hspace{0.5 cm}
   \resizebox{7 cm}{!}{\rotatebox{0}{\includegraphics{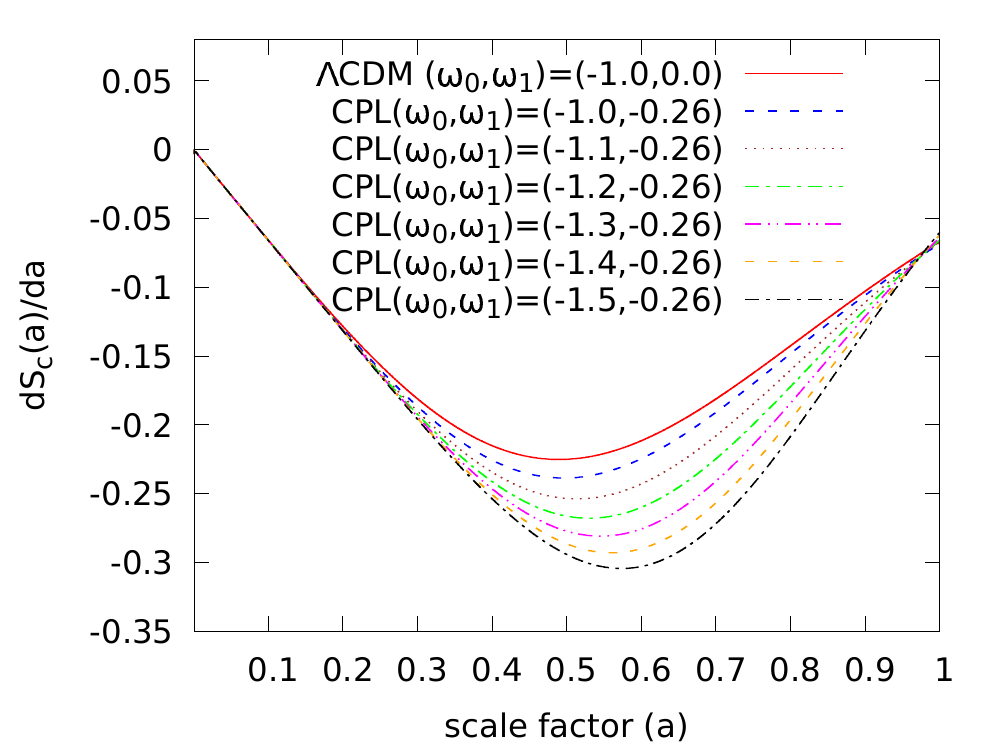}}}\\
   \caption{ The top left panel shows the evolution of the
     configuration entropy with scale factor for CPL parametrization
     with a fixed value of $\omega_0$ but different values of
     $\omega_1$. The top right panel show the entropy rate as a
     function of scale factor for the results shown in the top left
     panel. The bottom left panel shows the evolution of the
     configuration entropy with scale factor for CPL parametrization
     with a fixed value of $\omega_1$ but different values of
     $\omega_0$. The respective entropy rates are shown as a function
     of scale factor in the bottom right panel.
}
   \label{fig:cpl}
  \end{figure*}

\begin{figure*}

 \resizebox{7 cm}{!}{\rotatebox{0}{\includegraphics{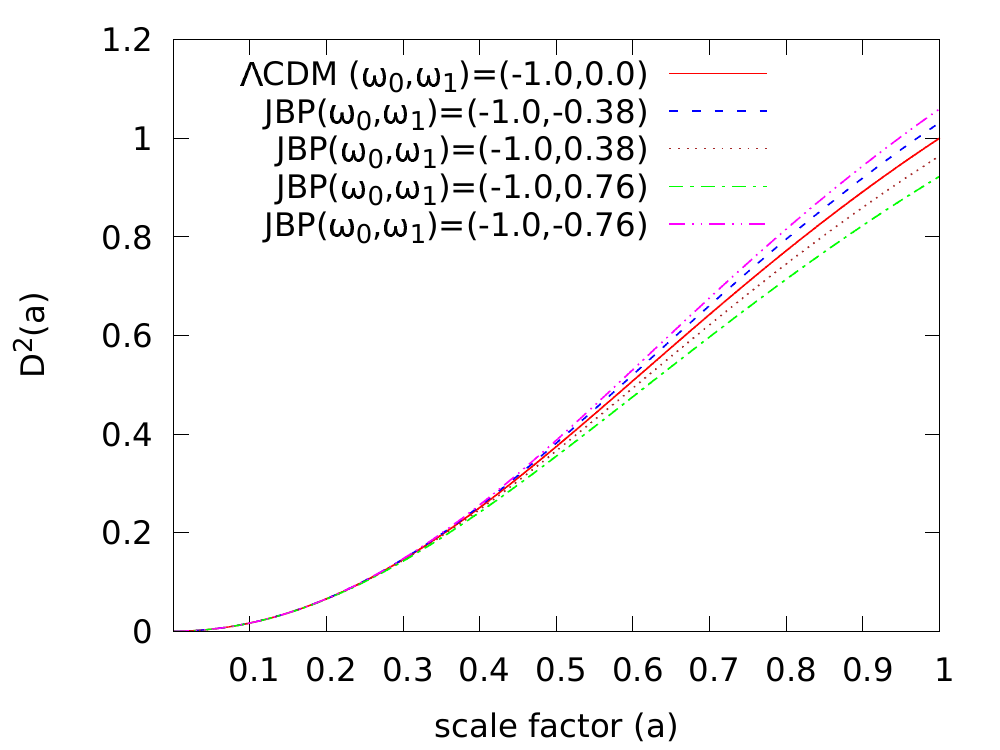}}}
  \hspace{0.5 cm}
  \resizebox{7 cm}{!}{\rotatebox{0}{\includegraphics{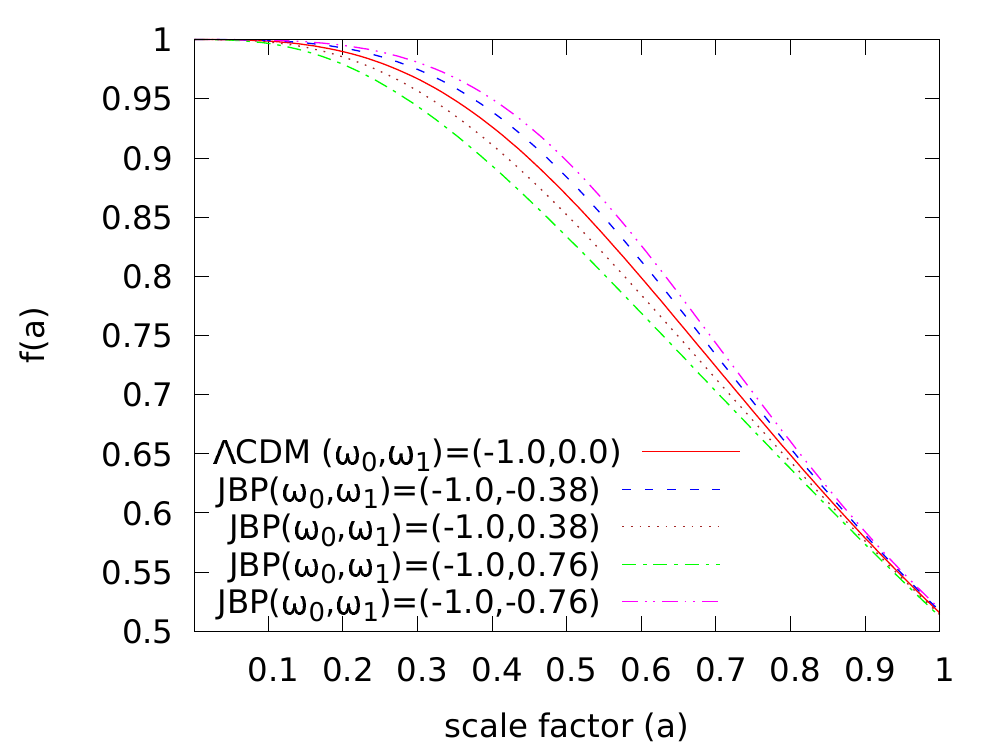}}}\\
   \vspace{-0.1 cm}
   \resizebox{7 cm}{!}{\rotatebox{0}{\includegraphics{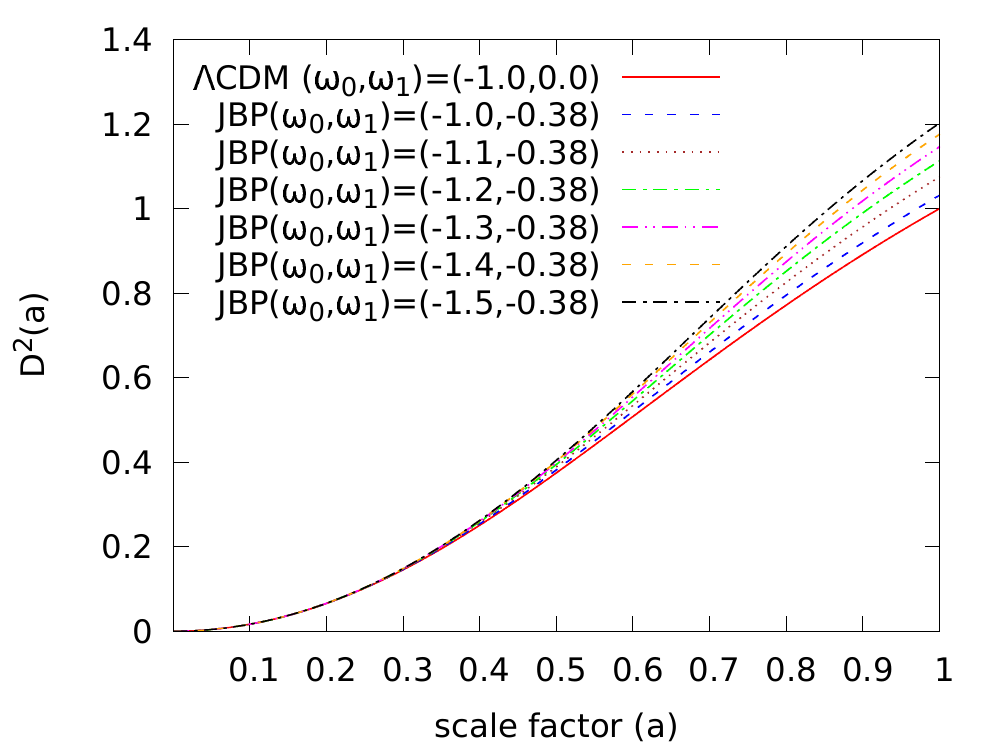}}}
   \hspace{0.5 cm}
   \resizebox{7 cm}{!}{\rotatebox{0}{\includegraphics{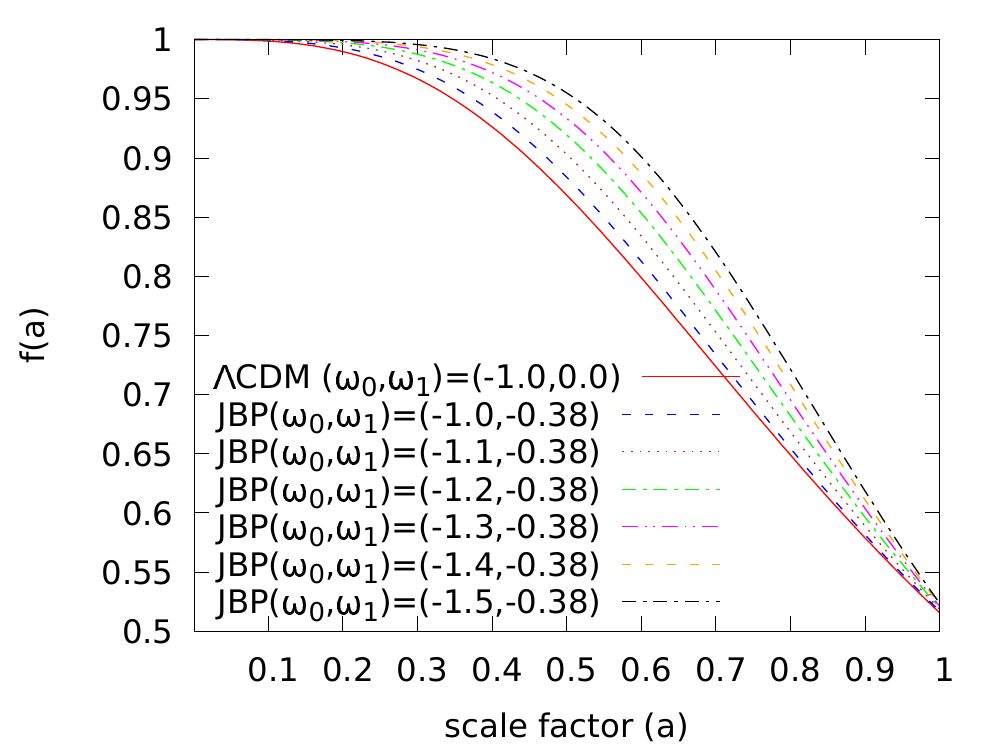}}}\\
   \caption{Same as \autoref{fig:cpl_fd} but for JBP
     parametrization. }
   \label{fig:jbp_fd}
  \end{figure*}

\begin{figure*}
   \resizebox{7 cm}{!}{\rotatebox{0}{\includegraphics{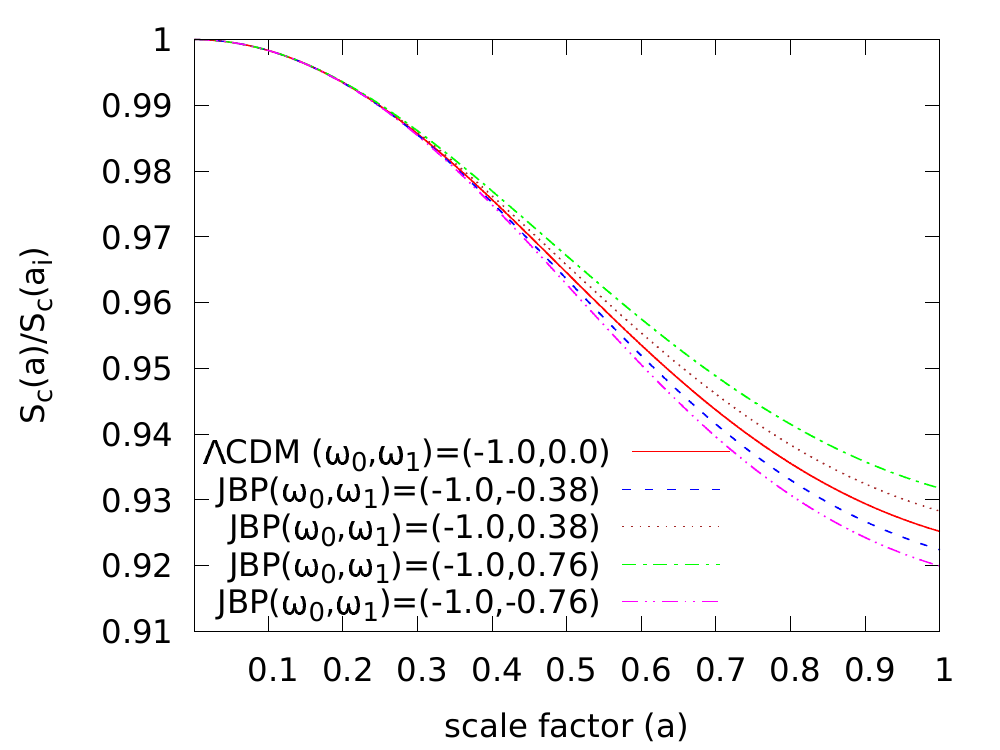}}}
  \hspace{0.5 cm}
  \resizebox{7 cm}{!}{\rotatebox{0}{\includegraphics{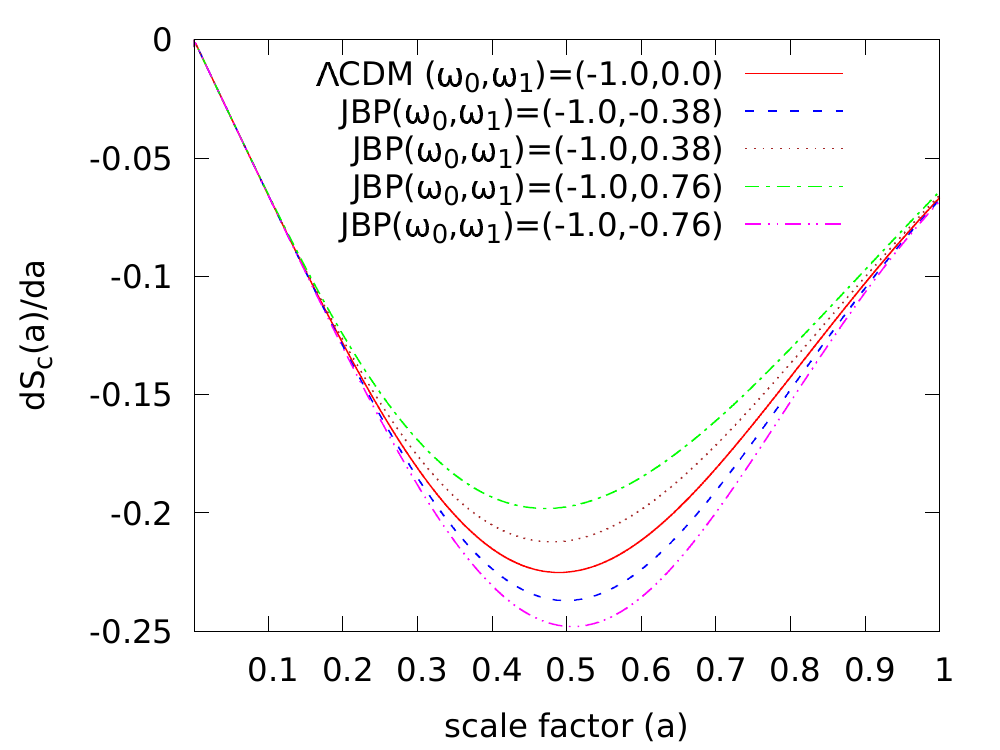}}}\\
   \vspace{-0.1 cm}
   \resizebox{7 cm}{!}{\rotatebox{0}{\includegraphics{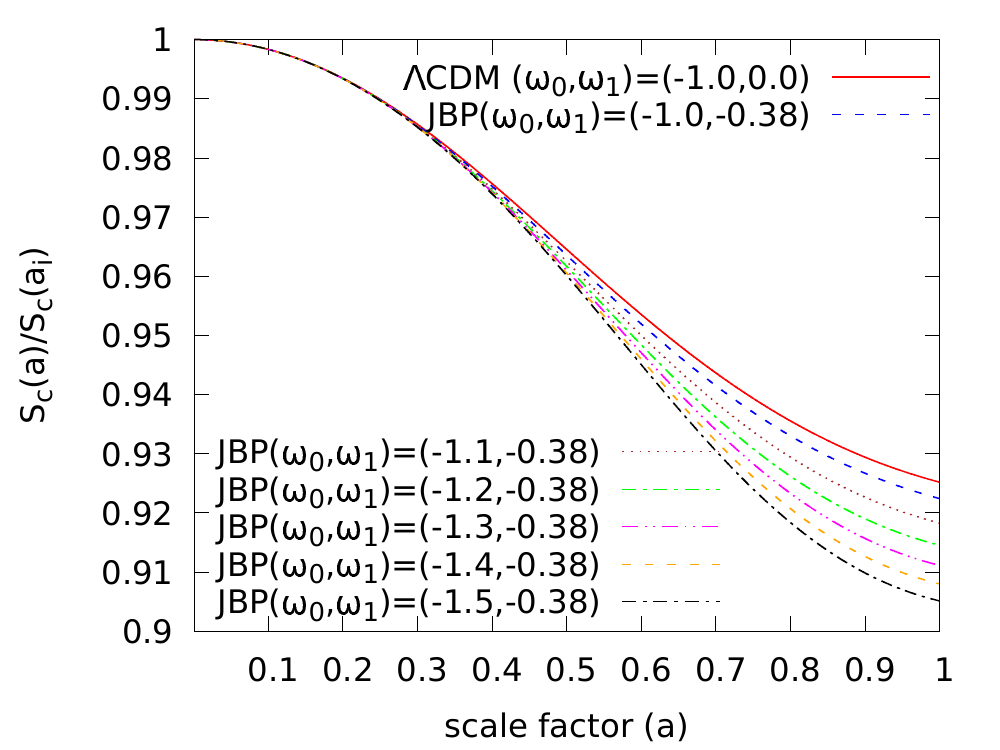}}}
   \hspace{0.5 cm}
   \resizebox{7 cm}{!}{\rotatebox{0}{\includegraphics{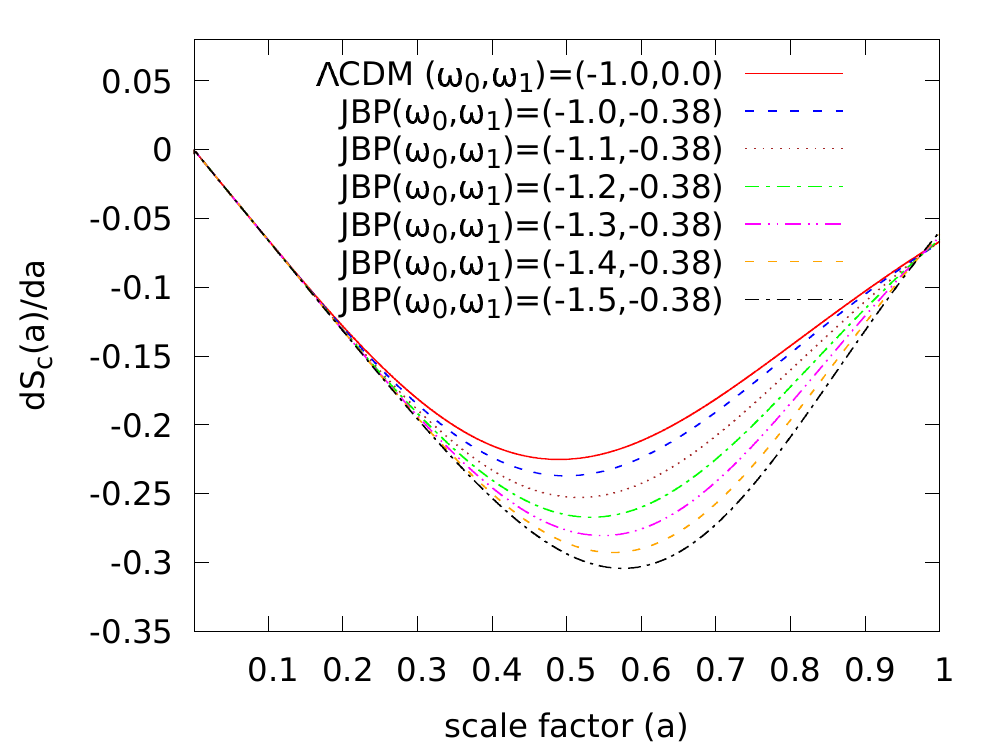}}}\\
   \caption{Same as \autoref{fig:cpl} but for JBP parametrization.}
   \label{fig:jbp}
   \end{figure*}

   \begin{figure*}
   \resizebox{7 cm}{!}{\rotatebox{0}{\includegraphics{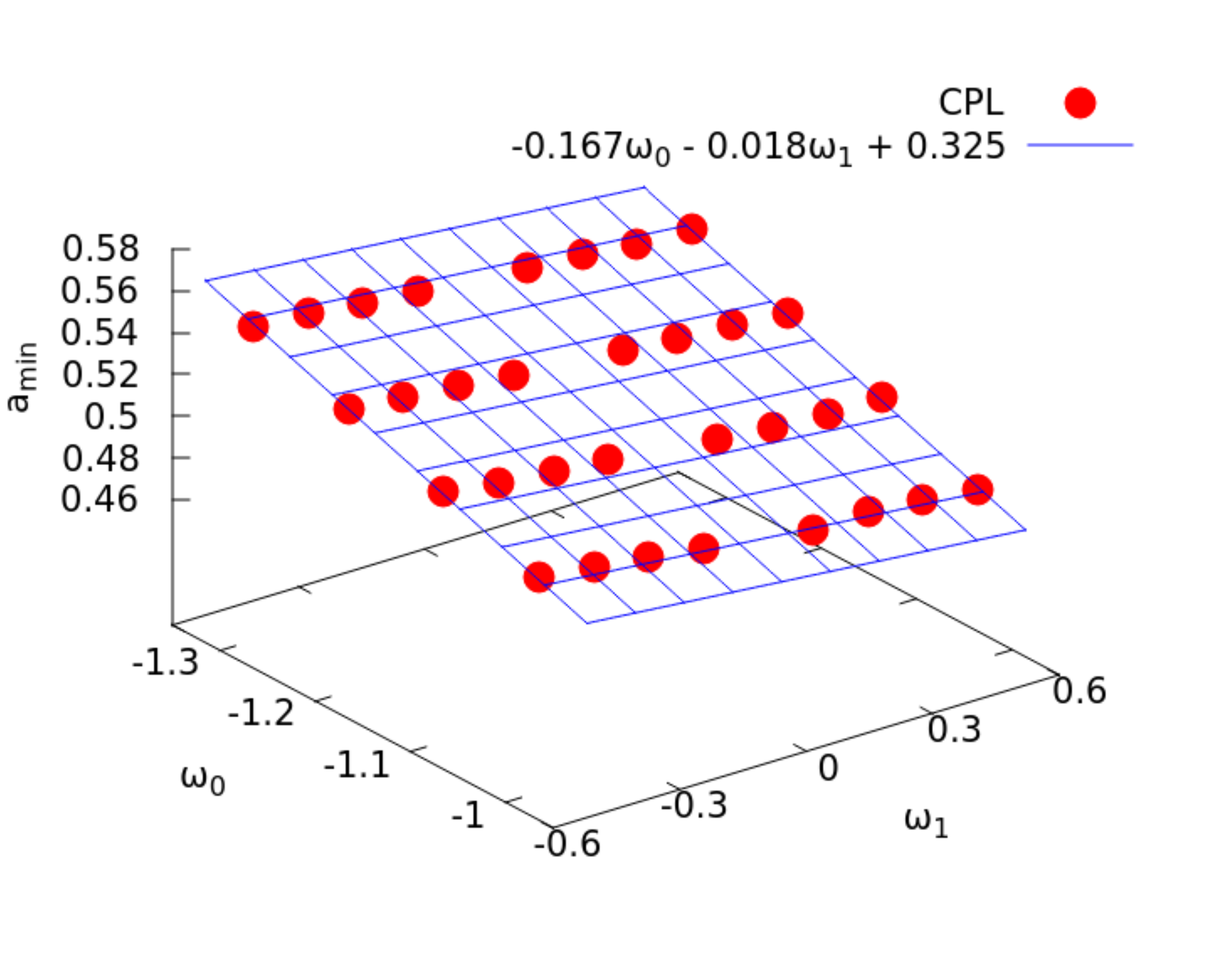}}}
   \hspace{0.5 cm}
   \resizebox{7 cm}{!}{\rotatebox{0}{\includegraphics{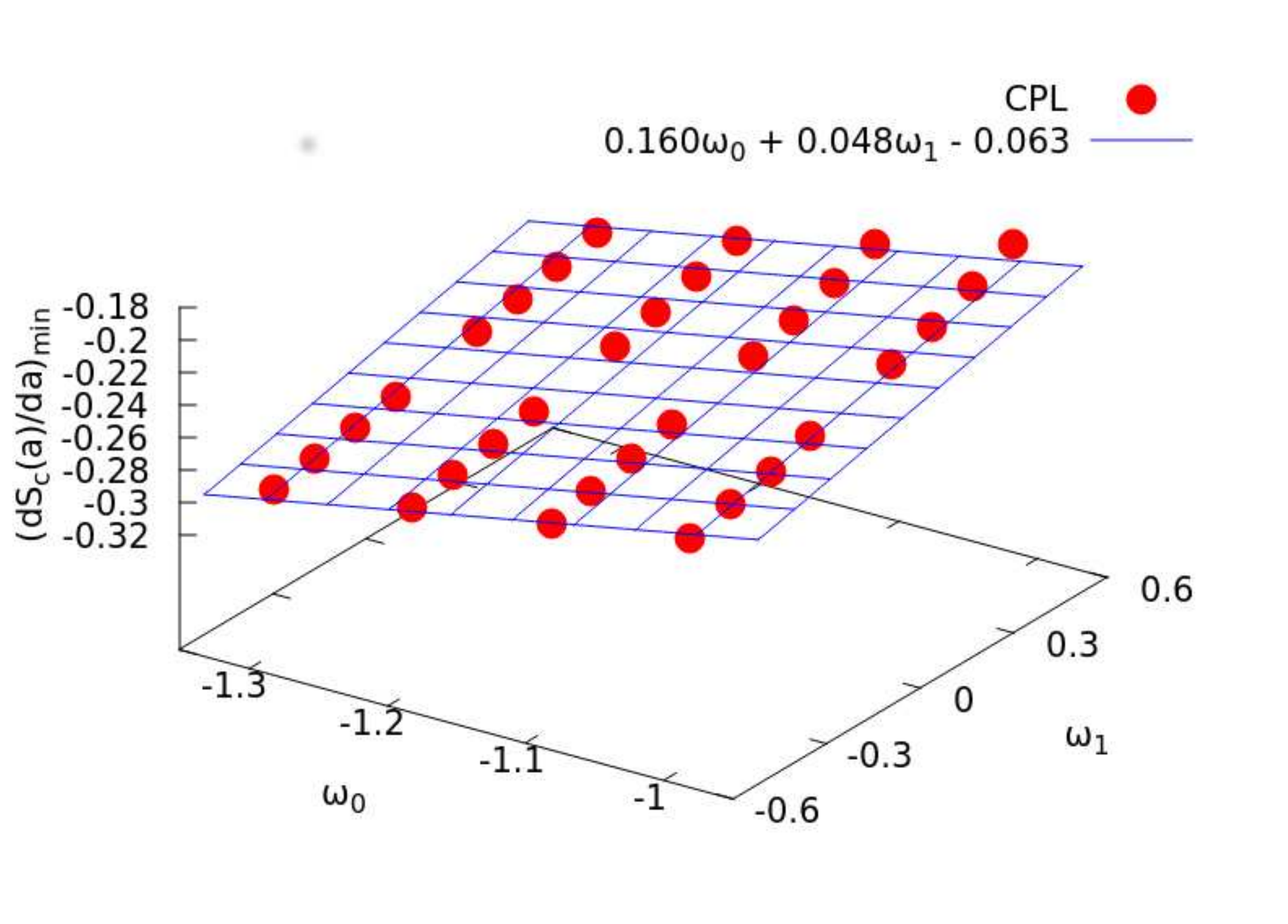}}}\\
   \vspace{-0.1 cm}
   \resizebox{7 cm}{!}{\rotatebox{0}{\includegraphics{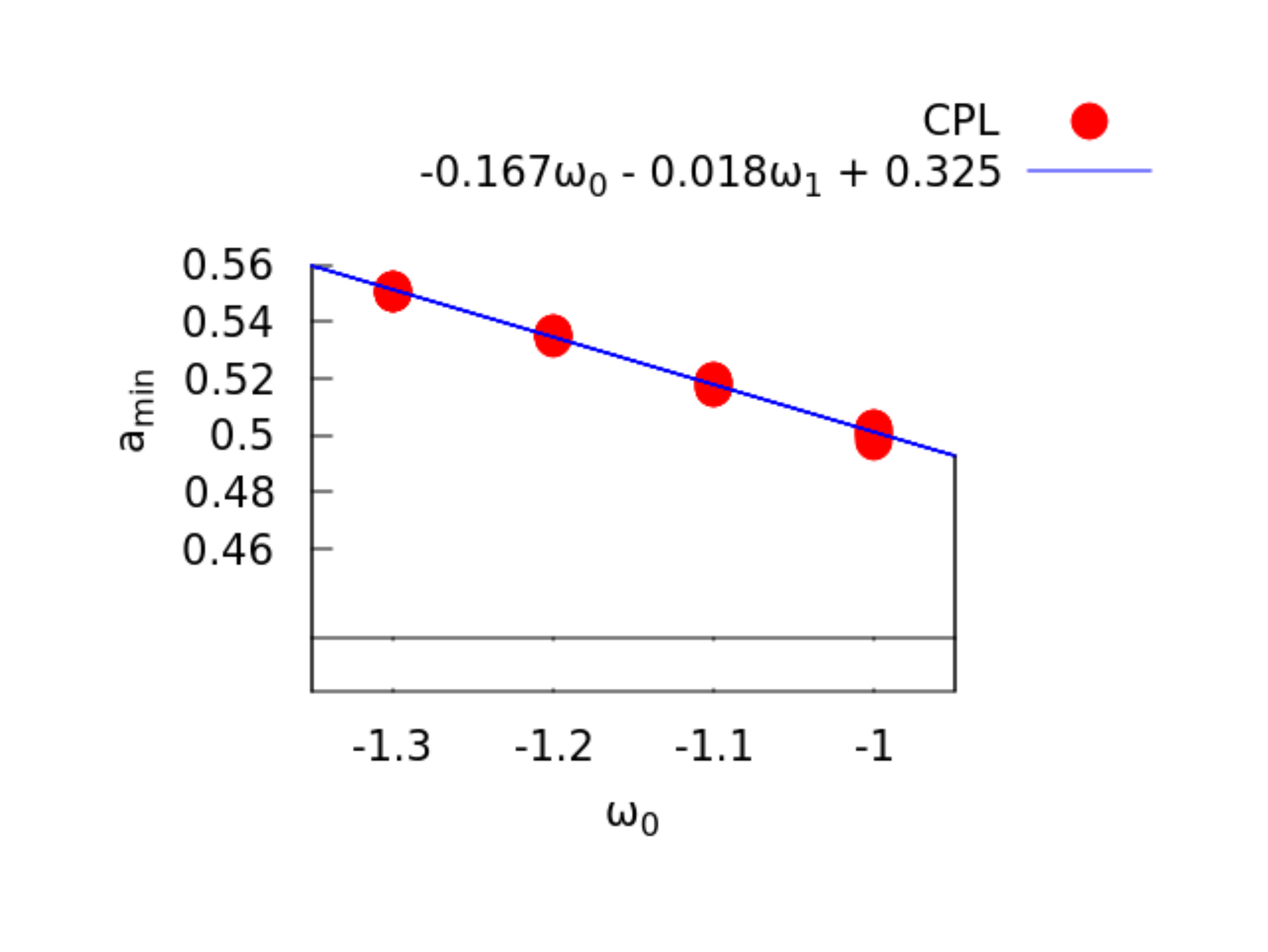}}}
   \hspace{0.5 cm}
   \resizebox{7 cm}{!}{\rotatebox{0}{\includegraphics{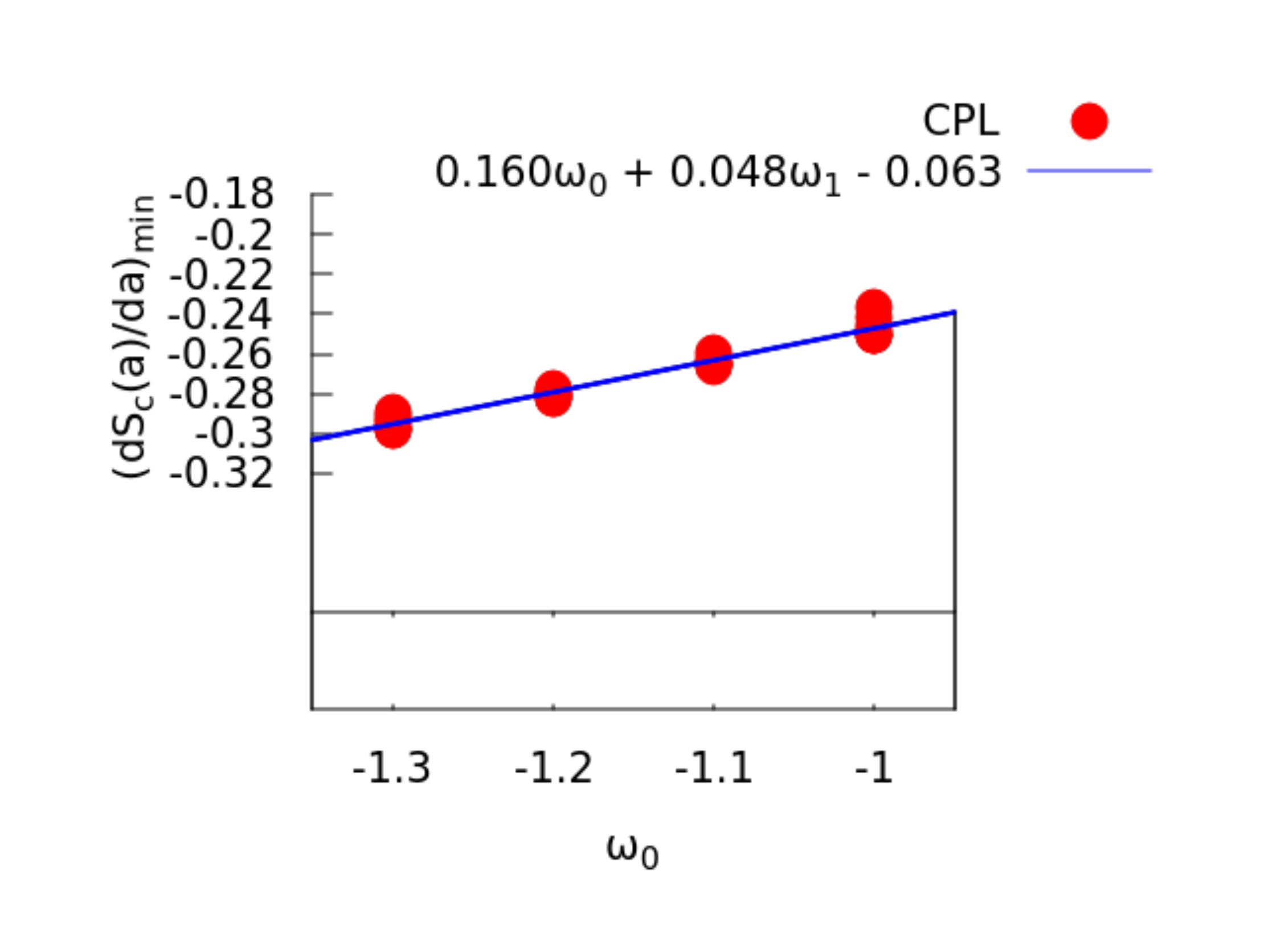}}}\\
   \hspace{0.5 cm}
   \resizebox{7 cm}{!}{\rotatebox{0}{\includegraphics{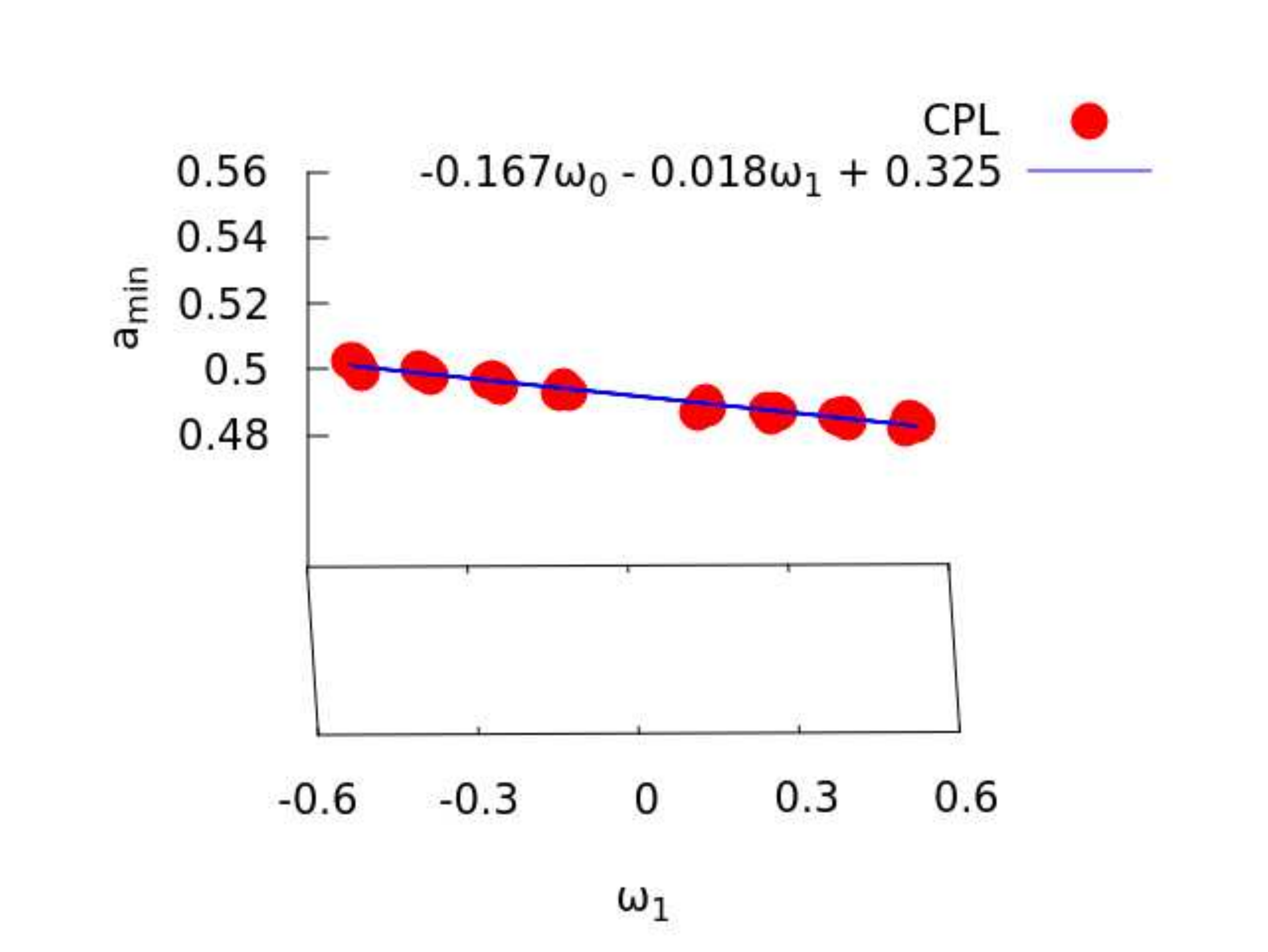}}}
   \hspace{0.5 cm}
   \resizebox{7 cm}{!}{\rotatebox{0}{\includegraphics{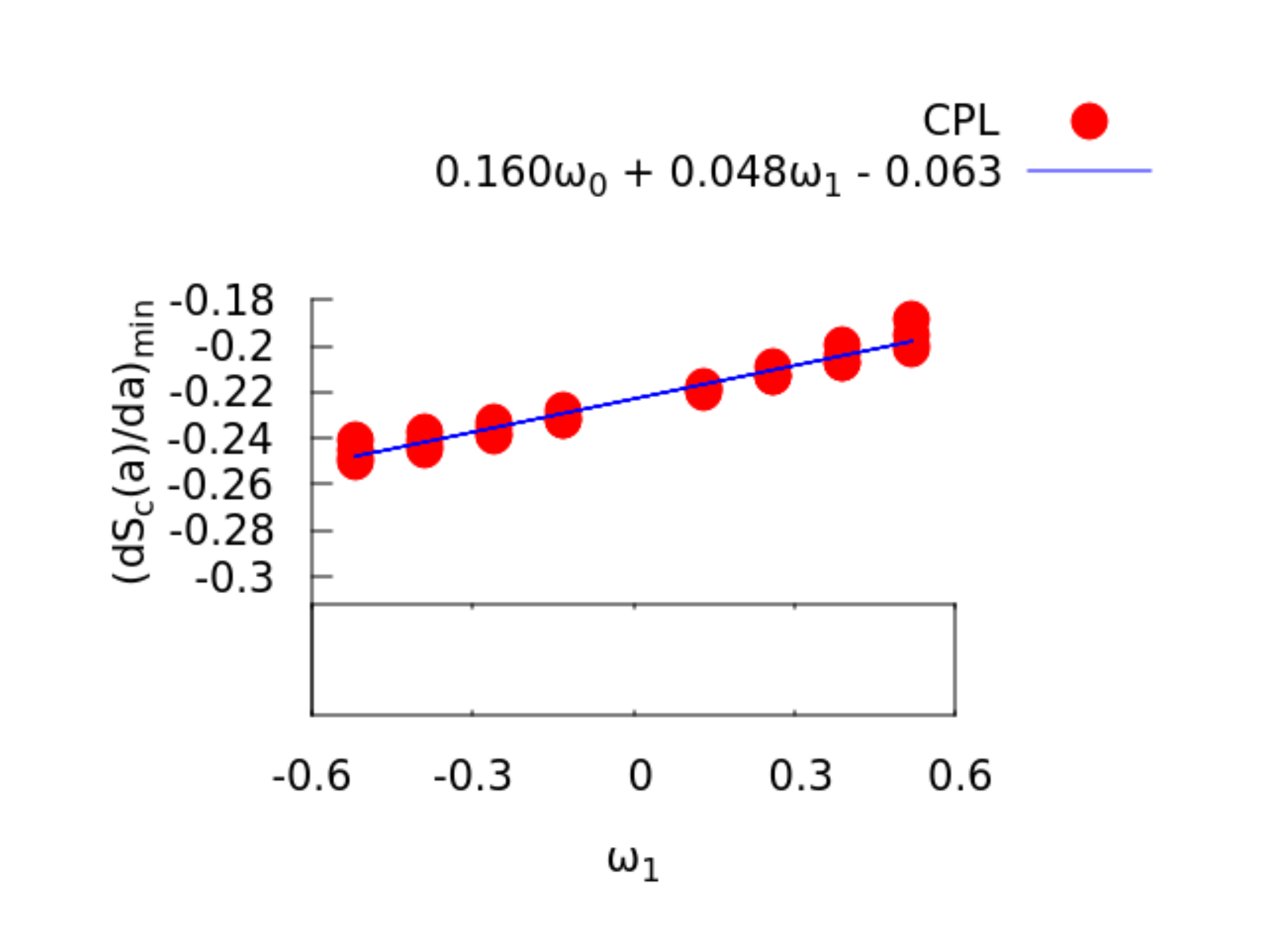}}}\\

   \caption{The top left and right panels respectively show
       the values of $a_{min}$ and $(\frac{dS_c(a)}{da})_{min}$ for
       different combinations of $\omega_0$ and $\omega_1$ in the CPL
       parametrization. The middle left and right panels respectively
       show the respective quantities as a function of $\omega_0$
       where we have stacked the results obtained for different
       $\omega_1$ values. The same quantities are shown as a function
       of $\omega_1$ by stacking the results obtained for different
       $\omega_0$ values in the lower left and right panels
       respectively. The best fit plane describing the result is
       also shown together in each panel.}
   \label{fig:cpl_plane}
   \end{figure*}

    \begin{figure*}
   \resizebox{7 cm}{!}{\rotatebox{0}{\includegraphics{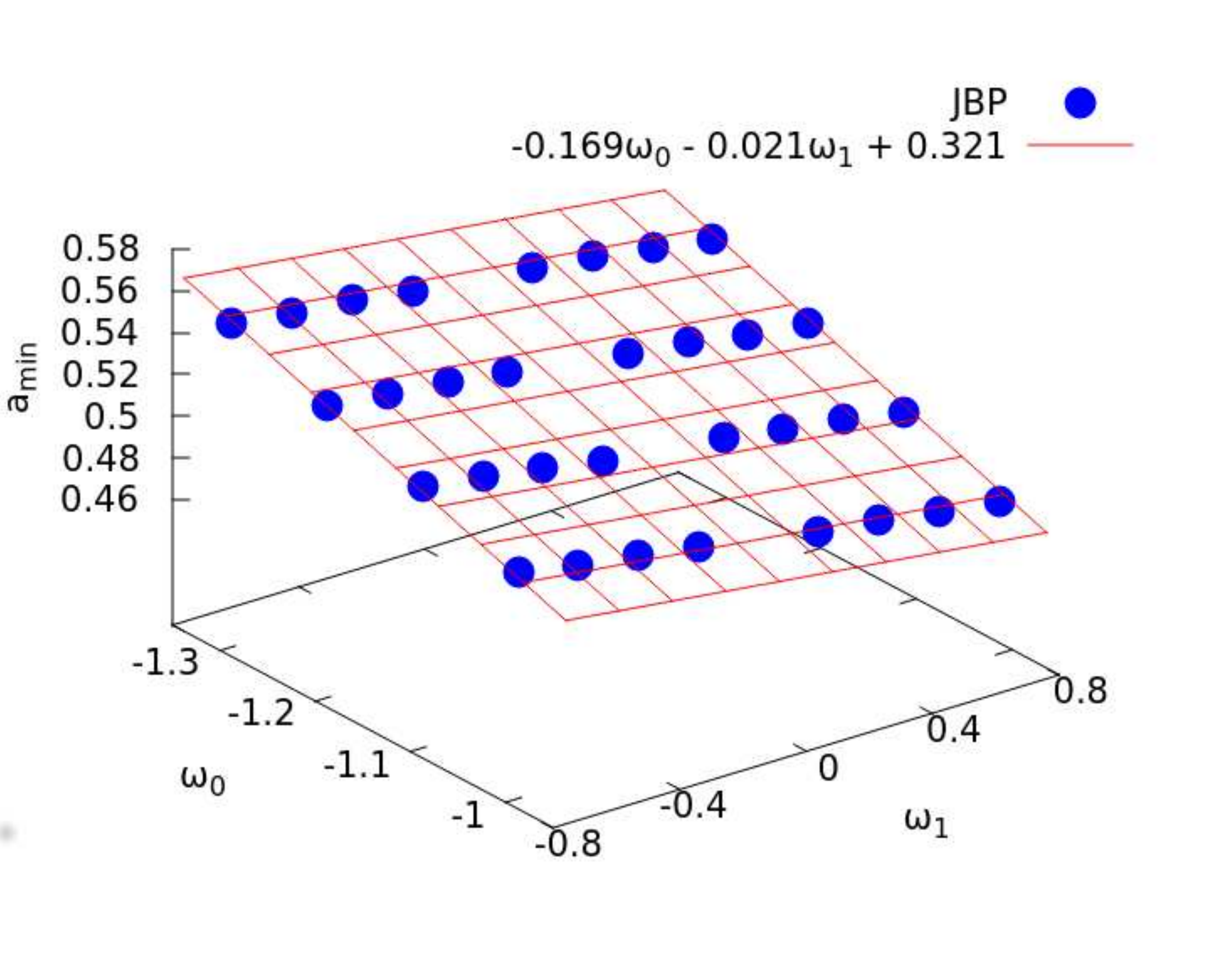}}}
   \hspace{0.5 cm}
   \resizebox{7 cm}{!}{\rotatebox{0}{\includegraphics{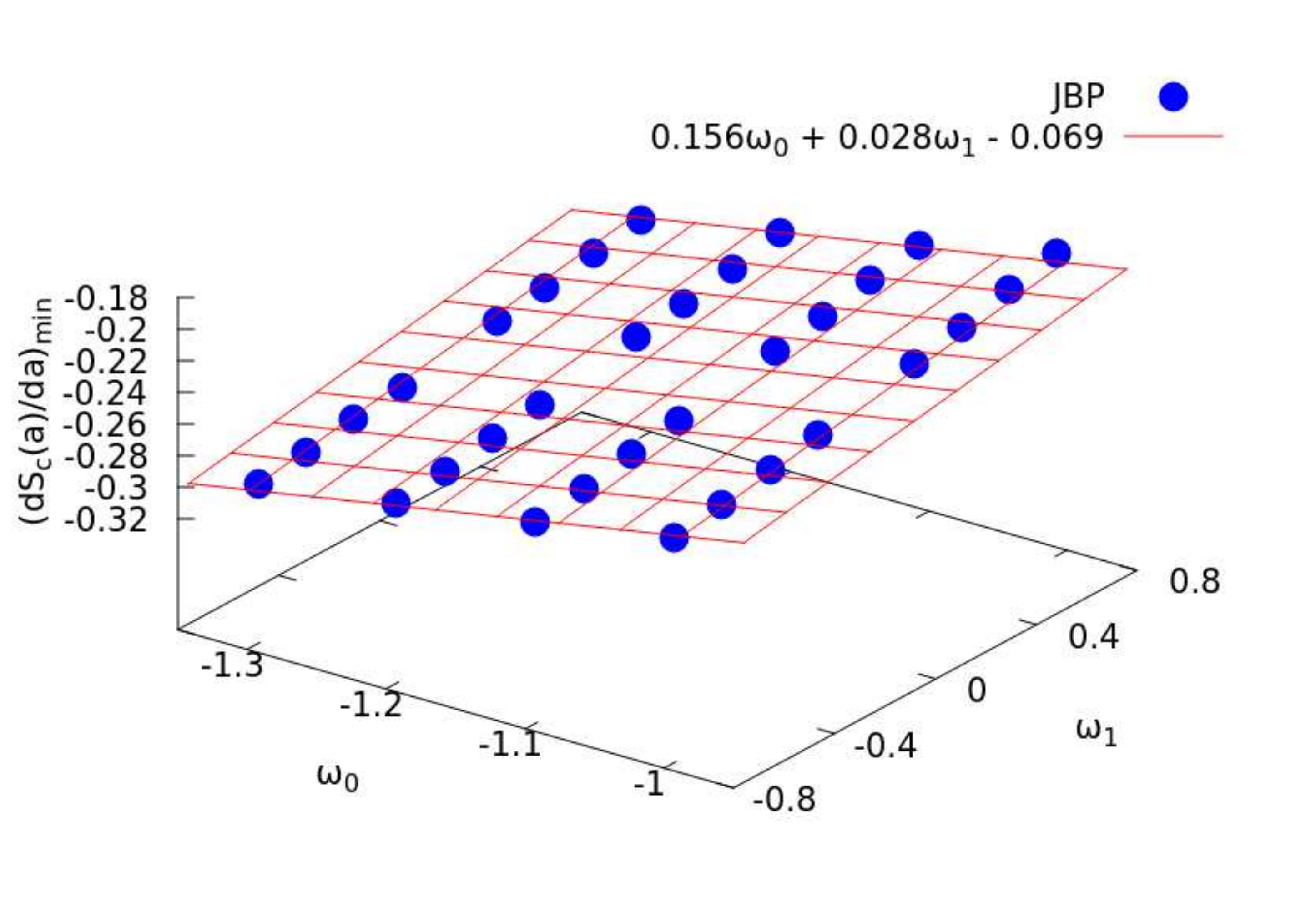}}}\\
   \vspace{-0.1 cm}
   \resizebox{7 cm}{!}{\rotatebox{0}{\includegraphics{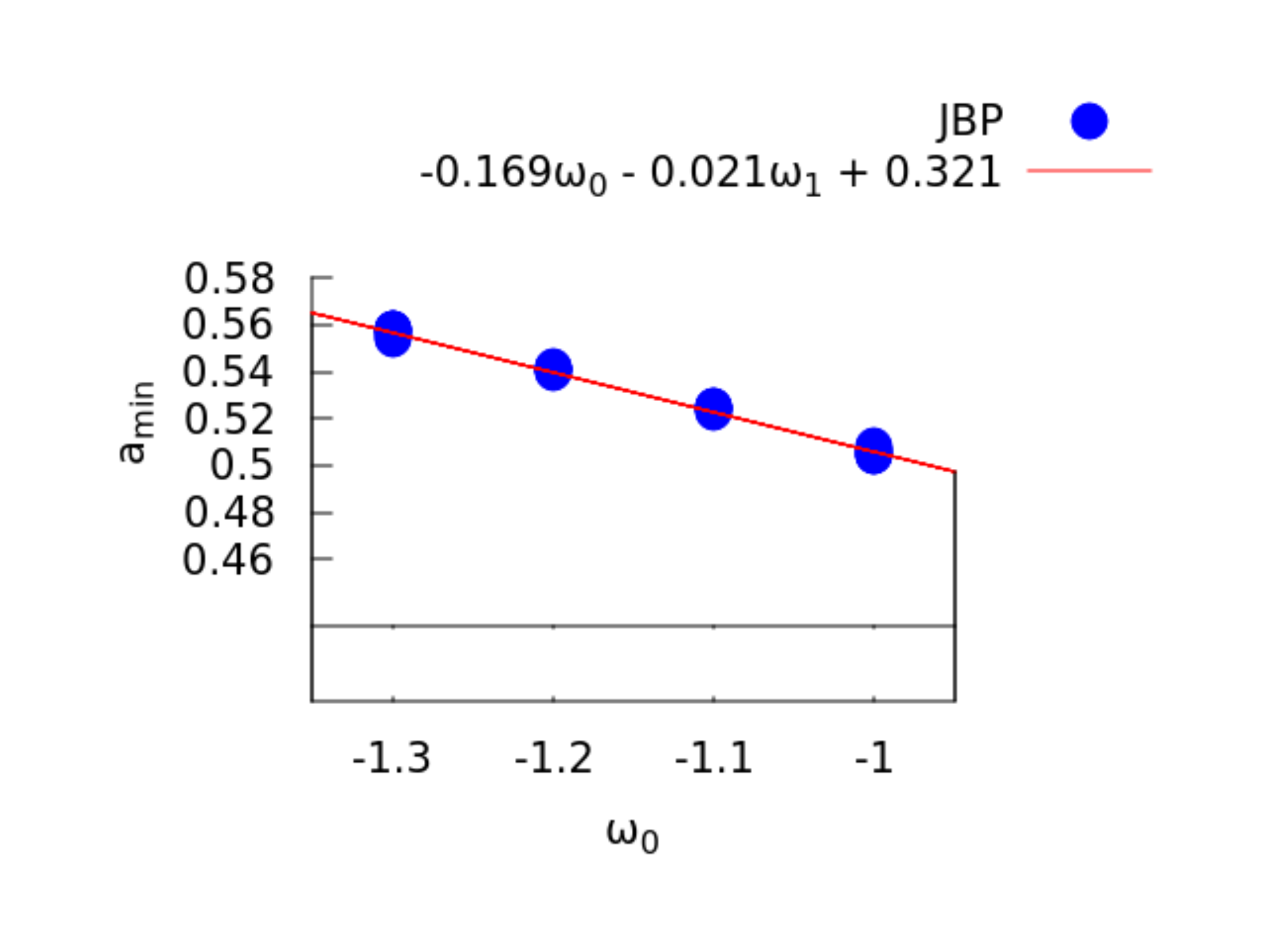}}}
   \hspace{0.5 cm}
   \resizebox{7 cm}{!}{\rotatebox{0}{\includegraphics{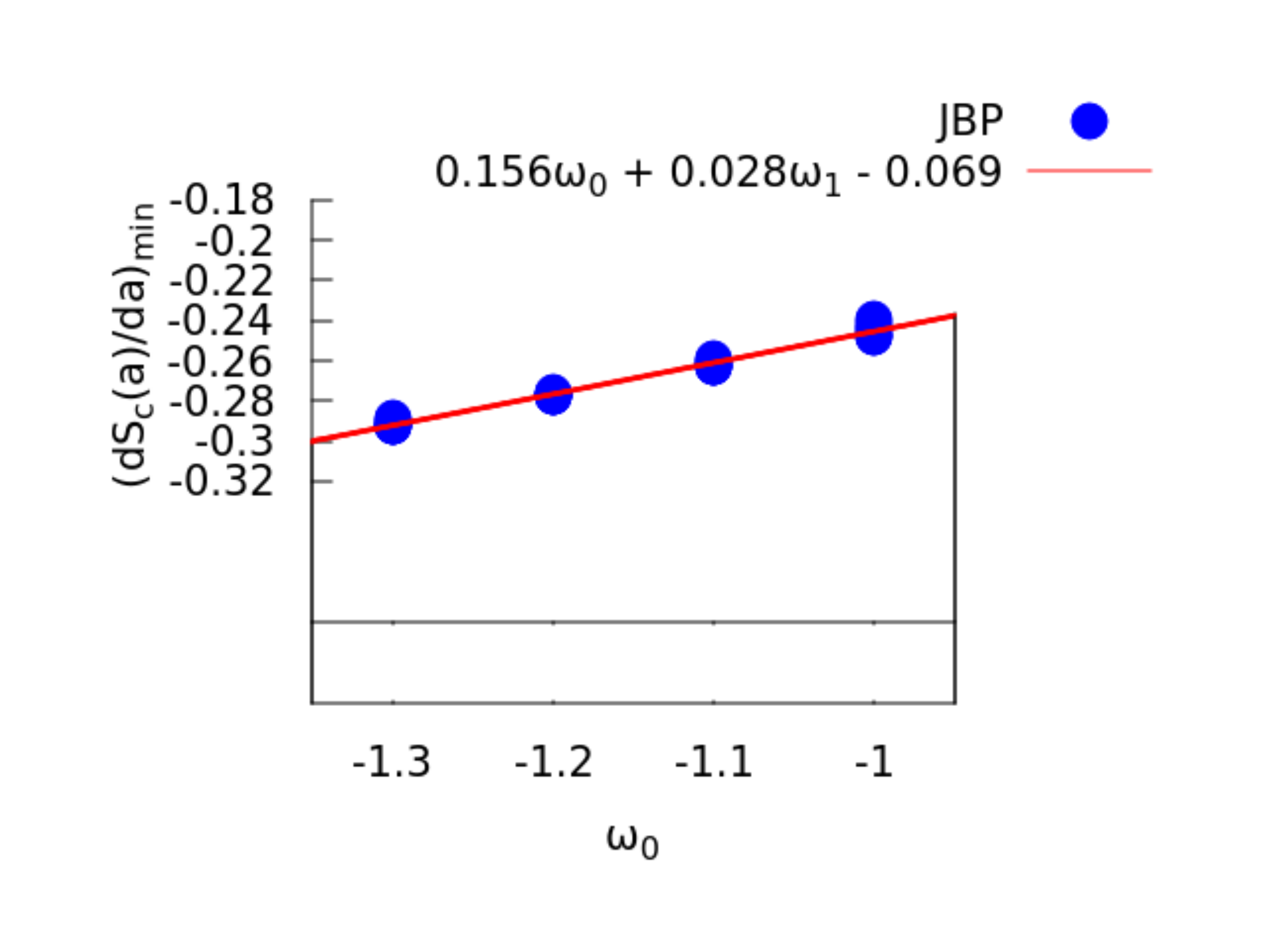}}}\\
   \hspace{0.5 cm}
   \resizebox{7 cm}{!}{\rotatebox{0}{\includegraphics{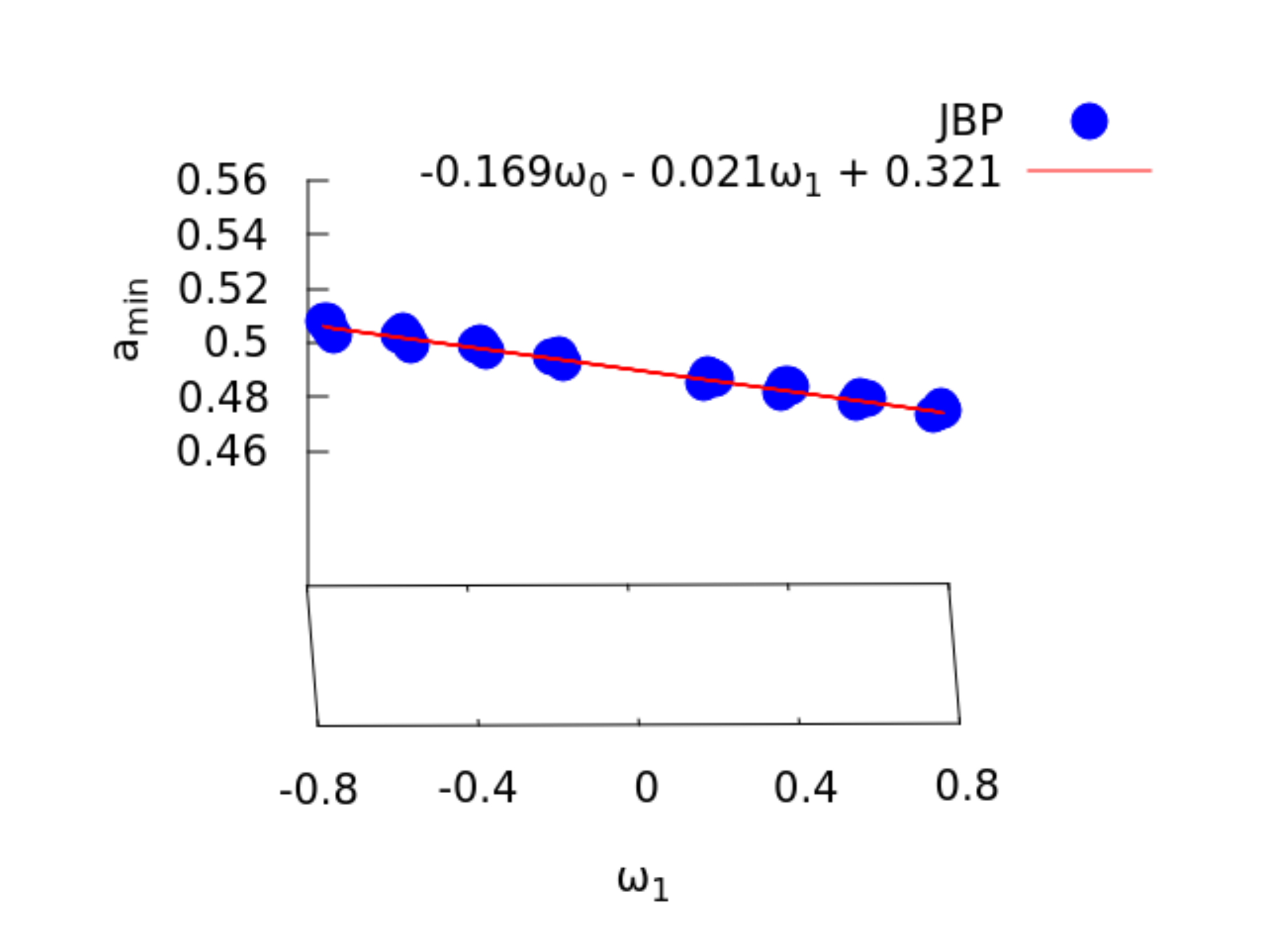}}}
   \hspace{0.5 cm}
   \resizebox{7 cm}{!}{\rotatebox{0}{\includegraphics{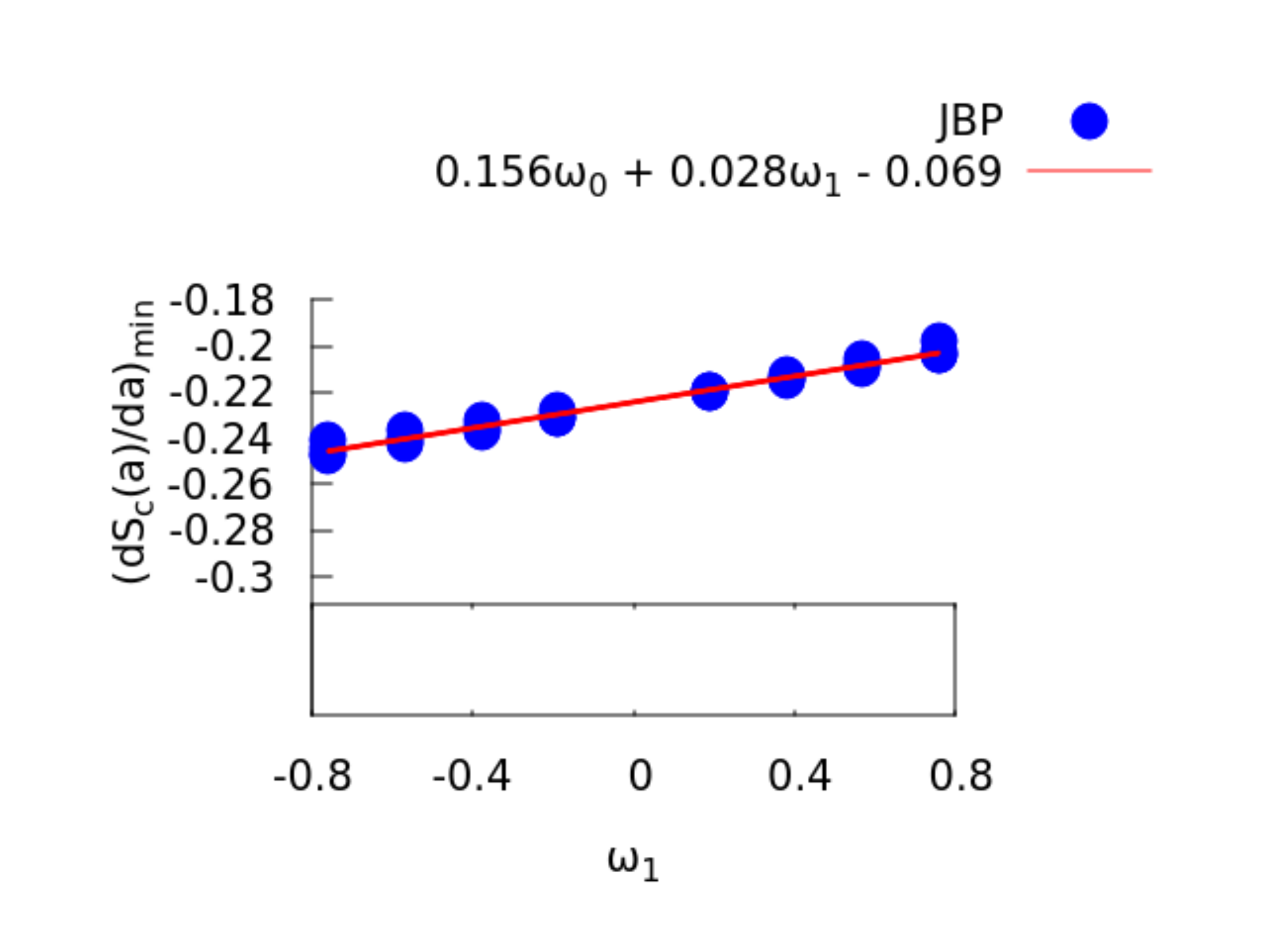}}}\\

   \caption{Same as \autoref{fig:cpl_plane} but for JBP parametrization.}
   \label{fig:jbp_plane}
  \end{figure*}

\subsection{Growth rate of density perturbations}

To explain the presence of structure in the Universe, it is presumed
that the inhomogeneities in the CMBR got amplified by the process of
gravitational instability over time. The growth of these primordial
density perturbations can be described by the the linear theory when
the density contrast, $\delta (\vec {x}, a) << 1$. In linear theory,
the time evolution of the density contrast is governed by the
following equation,
\begin{eqnarray}
	\frac{\partial^2 \delta (\vec {x}, t)}{\partial t^2} + 2 H(a) \frac{\partial \delta (\vec {x}, t)}{\partial t} - \frac{3}{2} \Omega_{m0} {H_0}^2 \frac {1}{a^3}  \delta (\vec {x}, t) = 0.
	\label{eq:nine}
\end{eqnarray}
Changing the variable of differentiation from $t$ to $a$ and introducing the deceleration parameter $q = - \frac {a \ddot a}{\dot a^2}$ we get \citep {linder4},
\begin{eqnarray}
	\frac {\partial^2 \delta (\vec {x}, a)}{\partial a^2} + \left( \frac {2 - q}{a}\right) \frac {\partial \delta (\vec {x}, a)}{\partial a} - \frac {3}{2} \frac {1}{a^2} \Omega_{m0} \delta (\vec {x}, a) = 0.
	\label{eq:ten}
\end{eqnarray}
The solution of \autoref{eq:nine} can be written as $\delta (\vec {x},
a) = d(a)\delta (\vec {x})$. Here $d(a)$ is the growing mode and
$\delta (\vec {x})$ is the initial density perturbation at the
comoving position $\vec {x}$. The change of variable
$D(a) = \frac{\delta (\vec {x}, a)}{\delta (\vec {x}, a_i)} = \frac{d(a)}{d(a_i)}$, where $a_i$ is some initial scale factor, leads to \citep {linder4},
\begin{eqnarray}
	\frac {d^2 D(a)}{da^2} + \frac{3}{2a} \left[1 - \frac{\omega(a)}{1 + X(a)}\right] \frac{dD(a)}{da} - \frac{3}{2} \frac {X(a)}{1 + X(a)}\frac{D(a)}{a^2} = 0.
	\label{eq:eleven}
\end{eqnarray}
Here 
\begin{eqnarray}
	X(a) = \frac {\Omega_{m0}}{1 - \Omega_{m0}} e^{- 3 \int_a^1 \omega (a^{\prime})\, d\log a^{\prime}}.   
	\label{eq:twelve}
\end{eqnarray}
$\Omega_{m0}$ is the present value of the mass density parameter and
$\omega(a)$ is the equation of state of dark energy. The time
dependence of dark energy is encoded in $\omega(a)$. We solve \autoref{eq:eleven} by the fourth order Runge-Kutta method. We normalize
$D(a_0) = 1$ in the $\Lambda$CDM model, where $a_0$ is the present
value of scale factor.

To find the dimensionless linear growth rate $f(a)$, we use 
\begin{eqnarray}
	f(a) = \left[\frac{\Omega_{m0} a^{-3}}{E^2(a)}\right]^{\gamma},
	\label{eq:thirteen}
\end{eqnarray}
where 
\begin{eqnarray}
	E^2(a) = \Omega_{m0} a^{-3} + (1 - \Omega_{m0})e^{3 \int_a^1 [1 + \omega (a^{\prime})]\,d \log a^{\prime}},
	\label{eq:fourteen}
\end{eqnarray}
and \citep {linder2}
\begin{eqnarray}
	\gamma = 0.55 + 0.02\, [1 + \omega (a = 0.5)].
	\label{eq:fifteen}
\end{eqnarray}

  \subsection{Different parametrizations of equation of state}
  Many different parametrizations of the equation of state of
  dynamical dark energy have been proposed in the literature which can
  be classified as one parameter and two parameter models depending on
  the number of parameter involved. We have considered a number of one
  parameter and two parameter models for our analysis. The
  parametrizations are briefly described in the following
  subsections. For each of the parametrizations, we use \autoref{eq:eleven}
  to find the evolution of $D(a)$ and then combine $D(a)$
  and $f(a)$ to find the evolution of entropy.

  \subsubsection{One parameter models}
We use a set of one parameter models for the dark energy equation of
state provided in \citet{yang}. The equation of states are given below
:

\begin{eqnarray}
\begin{array}{c c c}
	\mbox {Model 1} & \omega(a) = \omega_0 e^{(a - 1)} & \omega_0 = -1.367\\
	\mbox {Model 2} & \omega(a) = \omega_0 a [1 - \log(a)] & \omega_0 = -1.130\\
	\mbox {Model 3} & \omega(a) = \omega_0 a e^{(1 - a)} & \omega_0 = -1.163\\	
	\mbox {Model 4} & \omega(a) = \omega_0 a [1 + \sin (1 - a)] & \omega_0 = -1.244\\
	\mbox {Model 5} & \omega(a) = \omega_0 a [1 + \sin^{-1} (1 - a)] & \omega_0 = -1.213
	\label{tab:one}
\end{array}
\end{eqnarray}

All the parametrizations, except the first one approach zero as $a
\rightarrow 0$.  Each of these parametrizations have only one free
parameter $\omega_0$. The values of $\omega_0$ provided above are the
best fit values obtained by \citet {yang} using CMB + BAO + JLA + CC
data. We have also used these values besides the other values of
$\omega_0$ considered in our analysis.

\subsubsection{Two parameter models}
(i) CPL parametrization: The equation of state in the
Chevallier-Polarski-Linder parametrization \citep {chevallier,linder1}
is given as,
\begin{eqnarray}
	\omega(a) = \omega_0 + \omega_1 (1 - a)
	\label{seventeen}
\end{eqnarray}
which approaches $\omega_0 + \omega_1$ when $a$ approaches zero. The
equation of state changes with a constant slope $-\omega_1$. Apart
from the different possible combinations of $(\omega_0, \omega_1)$, we
also use in our analysis the best fit values $(\omega_0, \omega_1) =(-1.0, -0.26)$ obtained by \citet {tripathi} using SNIa + BAO + H(z)
data.

(ii) JBP parametrization: In Jassal-Bagla-Padmanabhan parametrization
\citep {jbp}, the equation of state is parametrized as
\begin{eqnarray}
	\omega(a) = \omega_0 + \omega_1 a (1 - a)
	\label{eq:eighteen}
\end{eqnarray}
This parametrization approaches $\omega_0$ as $a$ approaches zero. The
slope in this case is not constant but varies linearly with $a$. The
best fit values $(\omega_0, \omega_1) = (-1.0, -0.38)$ obtained by
\citet {tripathi} using SNIa + BAO + H(z) data is also used besides the
other possible combinations of $(\omega_0, \omega_1)$.

\begin{table*}{}
\caption{This shows $a_{min}$ and $(\frac{dS_c(a)}{da})_{min}$ as a
  function of the parameter/parameters of different
  parametrizations. The relations are obtained by fitting the
  numerical results obtained for each model.}
\label{tab:fits}
\begin{center}
  \begin{tabular}{|c|c|c|}
    \hline
    Model & $a_{min}$ & $(\frac{dS_c(a)}{da})_{min}$\\
    \hline
    Model 1 & $-0.151 \omega_0 + 0.331$ & $0.160 \omega_0 - 0.030$\\
    \hline
    Model 2 & $-0.158 \omega_0 + 0.337$ & $0.183 \omega_0 - 0.018$\\
    \hline
    Model 3 & $-0.152 \omega_0 + 0.345$ & $0.201 \omega_0 + 0.010$\\
    \hline
    Model 4 & $-0.154 \omega_0 + 0.343$ & $0.232 \omega_0 + 0.067$\\
    \hline
    Model 5 & $-0.153 \omega_0 + 0.344$ & $0.219 \omega_0 + 0.044$\\
    \hline
    CPL & $-0.167 \omega_0 - 0.018 \omega_1 + 0.325$ & $0.160 \omega_0 + 0.048 \omega_1 - 0.063$\\
    \hline
    JBP & $-0.169 \omega_0 - 0.021 \omega_1 + 0.321$ & $0.156 \omega_0 + 0.028 \omega_1 - 0.069$\\
    \hline
\end{tabular}
\end{center}
\end{table*}

\section{Results and Conclusions}
We show the square of the growing mode $D^2(a)$ and the dimensionless
linear growth rate $f(a)$ for different values of $\omega_0$ in Model
1 and in the $\Lambda$CDM model in the top left and top right panel of
\autoref{fig:one} respectively. In the middle left panel of
\autoref{fig:one}, we show the evolution of the configuration entropy
$S_c(a)$ with scale factor for different values of $\omega_0$ in Model
1. The result for the $\Lambda$CDM model is also shown together with
the Model 1 in the same panel. The derivative of the configuration
entropy $\frac{dS_c(a)}{da}$ as a function of scale factor for all the
cases are shown in the middle right panel of \autoref{fig:one}. The
configuration entropy dissipates due to the growth of
inhomogeneities. We observe that the entropy dissipation rate
initially increases with the increasing scale factor in all the
cases. But the derivative of the entropy dissipation rate eventually
changes sign at a specific scale factor. This scale factor $a_{min}$
corresponds to a minimum in the entropy rate. The entropy dissipation
rate slows down after the scale factor $a_{min}$. The magnitude of the
entropy rate $\frac{dS_c(a)}{da}$ at $a_{min}$ is directly related to
the growth rate of structures in a given model and it may be
  noted that the models with a higher growth rate exhibit a higher
  entropy dissipation rate. The value of $a_{min}$ indicates the
scale factor after which the dark energy plays an important role in
curbing the growth of structures in the Universe.  Both the value of
$a_{min}$ and the entropy rate $\frac{dS_c(a)}{da}$ at $a_{min}$ show
a systematic dependence on the parameter $\omega_0$ in the Model 1. We
calculate the values of $a_{min}$ and $(\frac{dS_c(a)}{da})_{min}$ in
Model 1 for different values of $\omega_0$. The bottom left and right
panels of \autoref{fig:one} respectively show $a_{min}$ and
$(\frac{dS_c(a)}{da})_{min}$ as a function of $\omega_{0}$ in Model
1. The best fit lines representing the numerical results
(\autoref{tab:fits}) are also plotted together in the two bottom
panels of \autoref{fig:one}. These results clearly indicate that the
monotonic dependence of $a_{min}$ and $(\frac{dS_c(a)}{da})_{min}$ on
$\omega_{0}$ in Model 1 can be used to constrain $\omega_{0}$ from the
observational study of the evolution of the configuration
entropy. Since there is only one free parameter in these models, one
can either use $a_{min}$ or $(\frac{dS_c(a)}{da})_{min}$ to constrain
the value of $\omega_0$ in Model 1.

  The results for Model 2, Model 3, Model 4 and Model 5 are shown in
  \autoref{fig:two}, \autoref{fig:three}, \autoref{fig:four} and
  \autoref{fig:five} respectively. We find that there exists a minimum
  in $\frac{dS_c(a)}{da}$ in all these models. The values of $a_{min}$
  and $(\frac{dS_c(a)}{da})_{min}$ albeit depend on the model and the
  specific value of $\omega_0$. These results suggest that one can
  describe the behaviour of $a_{min}$ and $(\frac{dS_c(a)}{da})_{min}$
  in terms of $\omega_0$ in each of these models. We find that both
  $a_{min}$ and $(\frac{dS_c(a)}{da})_{min}$ are linearly related to
  $\omega_0$. These linear relationships can be used to constrain the
  value of $\omega_0$ in the respective models. We find that the
  relationship between $a_{min}$ and $\omega_0$ are quite similar in
  all the models and hence it may not be very useful in distinguishing
  various one parameter models. Interestingly, the relationship
  between $(\frac{dS_c(a)}{da})_{min}$ and $\omega_0$ depends on the
  model (\autoref{tab:fits}). This arises due to the fact that the
  entropy dissipation rate is sensitive to the growth rate of
  structures and the equation of state has a direct influence on the
  growth rate of structures. So this relationship may be used to
  discern the model as well as constrain the value of $\omega_0$ in
  that model.  We also note that the location of the minimum of the
  entropy rate in the $\Lambda$CDM model deviates noticeably from the
  expectations for different values of $\omega_0$ in Model 3, Model 4
  and Model 5. So these models can be clearly distinguished from the
  $\Lambda$CDM model based on such an analysis.

  The results for the two-parameter models are shown in
  \autoref{fig:cpl} and \autoref{fig:jbp}. In an earlier work,
  \citet{das} show that the evolution of configuration entropy may
  help us to distinguish between different dark energy
  parametrizations. In the present work, we explore the possibility of
  constraining the parameters of a given parametrization by studying
  the evolution of the configuration entropy. We have considered the
  CPL and JBP parametrizations each of which has two parameters. We
  study how these parameters separately affect the evolution of the
  configuration entropy. The top left panel of \autoref{fig:cpl} shows
  the variation of entropy with scale factor for CPL parametrization
  by keeping $\omega_0$ fixed while varying $\omega_1$. We show the
  growing mode and the dimensionless linear growth rate for each set
  of EoS parameters in CPL and JBP parametrizations in
  \autoref{fig:cpl_fd} and \autoref{fig:jbp_fd} respectively. The
  results for the $\Lambda$CDM model is also shown together in each of
  the panels for comparison. The models with positive $\omega_1$ show
  less growth as compared to $\Lambda$CDM while the models with
  negative $\omega_1$ show higher growth as compared to
  $\Lambda$CDM. Consequently, the configuration entropy dissipates
  faster in the models with negative $\omega_1$. We show the
  configuration entropy rate in the top right panel of
  \autoref{fig:cpl}.  The derivative of the configuration entropy for
  the CPL parametrization also show the existence of a minimum. All
  the models show the minimum in entropy rate at almost the same scale
  factor. So the value of $a_{min}$ is less sensitive to the value of
  $\omega_1$. However, the magnitude of the entropy rate at $a_{min}$
  show a relatively stronger dependence on $\omega_1$. This is again
  related to the higher growth rate in the models with negative
  $\omega_1$.

In the two bottom panels of \autoref{fig:cpl}, we respectively show
the configuration entropy and its derivative as a function of scale
factor by keeping $\omega_1$ fixed and assuming different values for
$\omega_0$. We find that the location of the minimum of the entropy
rate systematically shifts towards higher values of scale factor with
decreasing values of $\omega_0$. The results clearly suggest that both
$a_{min}$ and $(\frac{dS_c(a)}{da})_{min}$ exhibit a relatively
stronger dependence on $\omega_0$ than $\omega_1$. 

The corresponding results for the JBP parametrization are shown in
different panels of \autoref{fig:jbp}. We observe a similar trend in
the behaviour of $a_{min}$ and $(\frac{dS_c(a)}{da})_{min}$ in case of
JBP parametrization. However these two quantities show a different
degree of dependence on $\omega_0$ and $\omega_1$ in the CPL and JBP
parametrizations.

In the top left and right panels of \autoref{fig:cpl_plane}, we
respectively plot the numerical values of $a_{min}$ and
$(\frac{dS_c(a)}{da})_{min}$ for different combinations of
$(\omega_0,\omega_1)$ in the CPL parametrization. We also show these
results as a function of $\omega_0$ by stacking them for all
$\omega_1$ in the two middle panels of \autoref{fig:cpl_plane}.
Similarly, these quantities are shown as a function of $\omega_1$ by
stacking the results for all $\omega_0$ in the two bottom panels of
this figure. The respective results for the JBP parametrization are
shown in the \autoref{fig:jbp_plane}. We also plot the best fitting
surface passing through the data points in all the panels. The
expressions for the best fitting planes are provided in
\autoref{tab:fits}. The results suggest that the dependence of
$a_{min}$ on $\omega_0$ and $\omega_1$ are quite similar in the CPL
and JBP parametrizations. We note that the dependence of
$(\frac{dS_c(a)}{da})_{min}$ on $\omega_0$ and $\omega_1$ are somewhat
different in the CPL and JBP parametrizations. The differences
primarily arise due to the differences in the growth history of
structures in the two parametrizations. For any given two parameter
model, the two best fitting equations for $a_{min}$ and
$(\frac{dS_c(a)}{da})_{min}$ (\autoref{tab:fits}) can be solved
together to determine $\omega_0$ and $\omega_1$ provided $a_{min}$ and
$(\frac{dS_c(a)}{da})_{min}$ are determined from observations.

The evolution of the configuration entropy is governed by the growing
mode and its derivative. So it may seem natural to directly use the
growth history of large scale structures to constrain the EoS
parameters \citep{linder4}. The possibility of using the growth rate
or growth index to distinguish different cosmological models have been
explored in the literature \citep{wang, linder2, gong}.  We can see in
Figures 1-6 and \autoref{fig:jbp_fd} that the growing mode and its
derivative are monotonic functions of scale factor in all the models
across different parametrizations considered in this work.  It would
be difficult to constrain the EoS parameters from these quantities
given their monotonic behaviour. On the other hand, the entropy rate
exhibits a distinct minimum and the location and amplitude of the
minimum are sensitive to the EoS parameters and the
parametrizations. The amplitude and the location of the minimum are
decided by the relative dominance of the dark energy and its effect on
the growth history of large scale structures. It may be noted that the
configuration entropy rate depends on a specific combination of the
growing mode, its derivative and the scale factor (the 3rd term in
\autoref{eq:six}). This specific combination is responsible for the
distinct minimum observed in the derivative of the configuration
entropy and we propose to use the location and amplitude of the
minimum as a probe of the EoS parameters.

We would also like to point out here that the future 21 cm
observations may enable us to measure the neutral Hydrogen
distributions at different redshifts. This would then allow us to
directly measure the configuration entropy without measuring the
growing mode and its derivative. If the evolution of the HI bias
\citep{bagla, guhasarkar, padmanabhan, sarkar_bharadwaj} can be
measured from these observations, the method presented in this work
can be then applied to such data sets as an independent and
alternative method to constrain the EoS parameters.

In this work, we propose an alternative scheme to constrain the
parameters of the dynamical dark energy models by studying the time
evolution of the configuration entropy in the Universe. In future, a
combined analysis of the present generation redshift surveys
(e.g. SDSS), the future generation surveys (e.g. Euclid) and the
future 21 cm experiments (e.g. SKA) may allow us to probe the
evolution of the configuration entropy in the Universe. The method
presented in this work would then allow us to constrain the equation
of state parameter/parameters for any given parametrization of the
dark energy.

\section {Acknowledgement}
The authors thank an anonymous reviewer for useful comments and
suggestions which helped to improve the manuscript. BP acknowledges
financial support from the Science and Engineering Research Board
(SERB), Department of Science \& Technology (DST), Government of India
through the project EMR/2015/001037. BP would also like to acknowledge
IUCAA, Pune for providing support through the associateship programme.

\bsp	
\label{lastpage}

\begin{thebibliography}{99}

\bibitem[Armendariz-Picon et al.(2001)]{armendariz} Armendariz-Picon,
  C., Mukhanov, V., \& Steinhardt, P.~J.\ 2001, \prd, 63, 103510
  
\bibitem[Amendola \& Tsujikawa(2010)]{de2010} Amendola, L. \& Tsujikawa, S.
  \ 2010 Dark Energy: Theory and Observation, Cambridge University Press

\bibitem[\protect\citeauthoryear{Bagla, Khandai \&
    Datta}{2010}]{bagla} Bagla J.~S., Khandai N., Datta K.~K., 2010,
  MNRAS, 407, 567

\bibitem[Brans \& Dicke(1961)]{bransdicke} Brans, C. \& Dicke,
  R.~H.\ 1961, Physical Review, 124, 925

\bibitem[Buchdahl(1970)]{buchdahl} Buchdahl, H.~A.\ 1970, \mnras, 150,
  1
\bibitem[Buchert(2000)]{buchert2k} Buchert, T.\ 2000, General
  Relativity and Gravitation, 32, 105

\bibitem[Caldwell et al.(1998)]{caldwell} Caldwell, R.~R., Dave, R.,
  \& Steinhardt, P.~J.\ 1998, Physical Review Letters, 80, 1582

\bibitem[Chevallier \& Polarski(2001)]{chevallier} Chevallier, M., 
  \& Polarski, D.\ 2001, International Journal of Modern Physics D, 10, 213

\bibitem[Copeland et al.(2006)]{copeland} Copeland, E.~J., Sami, M.,
  \& Tsujikawa, S.\ 2006, International Journal of Modern Physics D,
  15, 1753

\bibitem[\protect\citeauthoryear{Das \& Pandey}{2019}]{das} Das, B.,
  Pandey, B., 2019, MNRAS, 482, 3219

\bibitem[Easson et al.(2011)]{easson} Easson, D.~A., Frampton, P.~H.,
  \& Smoot, G.~F.\ 2011, Physics Letters B, 696, 273

\bibitem[Gong et al. (2009)]{gong} Gong Y., Ishak M., Wang A., 2009,
  \prd, 80, 023002
  
\bibitem[\protect\citeauthoryear{Guha Sarkar, et
    al.}{2012}]{guhasarkar} Guha Sarkar T., Mitra S., Majumdar S.,
  Choudhury T.~R., 2012, MNRAS, 421, 3570

\bibitem[Hunt \& Sarkar(2010)]{hunt} Hunt, P. \& Sarkar, S. \ 2010,
  \mnras, 401, 547

\bibitem[Jassal et al.(2005)]{jbp} Jassal, H.~K., Bagla, J.~S., 
  \& Padmanabhan, T.\ 2005, \mnras, 356, L11 

 \bibitem[Linder(2003)]{linder1} Linder, E.~V.\ 2003, Physical Review Letters, 
	90, 091301  
 
\bibitem[Linder(2005)]{linder2} Linder, E.~V.\ 2005, \prd, 72, 043529

\bibitem[Linder \& Jenkins(2003)]{linder4} Linder, E.~V., \& Jenkins, A.
	\ 2003, \mnras, 346, 573 

\bibitem[Milton(2003)]{milton} Milton, K.~A.\ 2003, Gravitation and Cosmology,
	9, 66 

\bibitem[\protect\citeauthoryear{Padmanabhan, Choudhury \&
    Refregier}{2015}]{padmanabhan} Padmanabhan H., Choudhury T.~R.,
  Refregier A., 2015, MNRAS, 447, 3745
        
\bibitem[Padmanabhan(2017)]{paddy} Padmanabhan, T.\ 2017, Comptes Rendus
        Physique, 18, 275,

\bibitem[Padmanabhan \& Padmanabhan(2017)]{paddyhamsa}
  Padmanabhan, T., \& Padmanabhan, H.\ 2017, Physics Letters B, 773,
  81
\bibitem[Pandey(2017)]{pandey1} Pandey, B.\ 2017, \mnrasl, 471, L77

\bibitem[Pandey(2019)]{pandey3} Pandey, B.\ 2019, \mnrasl, 485, L73

\bibitem[\protect\citeauthoryear{Pandey \& Das}{2019}]{pandey2} Pandey B.,
  Das, B.\ 2019, \mnrasl, 485, L43

\bibitem[Pav{\'o}n \& Radicella(2013)]{pavon1} Pav{\'o}n, D., \&
  Radicella, N.\ 2013, General Relativity and Gravitation, 45, 63

\bibitem[Perlmutter et al.(1999)]{perlmutter} Perlmutter, S.,
  Aldering, G., Goldhaber, G., et al.\ 1999, \apj, 517, 565

\bibitem[Radicella \& Pav{\'o}n(2012)]{radicella} Radicella, N., \&
  Pav{\'o}n, D.\ 2012, General Relativity and Gravitation, 44, 685

\bibitem[Ratra \& Peebles(1988)]{ratra} Ratra, B., \& 
  Peebles, P.~J.~E.\ 1988, \prd, 37, 3406 

\bibitem[Riess et al.(1998)]{riess} Riess, A.~G.,
  Filippenko, A.~V., Challis, P., et al.\ 1998, \aj, 116, 1009

\bibitem[\protect\citeauthoryear{Sarkar, Bharadwaj \&
    Anathpindika}{2016}]{sarkar_bharadwaj} Sarkar D., Bharadwaj S.,
  Anathpindika S., 2016, MNRAS, 460, 4310

\bibitem[Shannon(1948)]{shannon48} Shannon, C. E. \ 1948, Bell
System Technical Journal, 27, 379-423, 623-656

\bibitem[Tripathi et al.(2017)]{tripathi} Tripathi, A., Sangwan, A., 
	\& Jassal, H.~K.\ 2017, \jcap, 6, 012

\bibitem[Tomita(2001)]{tomita01} Tomita, K. \ 2001, \mnras, 326, 287

\bibitem[Wang \& Steinhardt (1998)]{wang} Wang L., Steinhardt P.~J., 1998,
  \apj, 508, 483
  
\bibitem[Yang et al.(2018)]{yang} Yang W., Pan S., Di Valentino E.,
  Saridakis E.~N., Chakraborty S.\ 2018, \prd, 99, 043543

\end{thebibliography}
\end{document}